\documentclass[prd,aps,showpacs,%11pt,%showpacs,
nofootinbib,%
notitlepage,
twocolumn%,superscriptaddress,
]{revtex4-2}

\usepackage{graphicx}
\usepackage{amsmath}
\usepackage{amssymb}
\usepackage{hyperref} 
\usepackage{upgreek}
\usepackage{tikz-feynman,contour}

\usepackage{color}
\usepackage[normalem]%
{ulem}  % \sout{old text} for strikeout

\newcommand{\1}{\color{black}}
\newcommand{\0}{\color{black}}

\newcommand{\be}{\begin{equation}}
\newcommand{\ee}{\end{equation}}
\newcommand{\bea}{\begin{eqnarray}}
\newcommand{\eea}{\end{eqnarray}}

\newcommand{\9}[1]{\breve{#1}}

\newcommand{\F}{\mathcal{F}}
\newcommand{\J}{\mathcal{J}}
\newcommand{\M}{\mathcal{M}}
\newcommand{\Q}{\mathcal{Q}}
\newcommand{\m}{W}

\begin{document}

\title{Superconnections in AdS/QCD and\\ the hadronic light-by-light contribution to the muon $g-2$}

\author{Josef Leutgeb}
\author{Jonas Mager}
\author{Anton Rebhan}
\affiliation{Institut f\"ur Theoretische Physik, Technische Universit\"at Wien,
        Wiedner Hauptstrasse 8-10, A-1040 Vienna, Austria}

\date{\today}

\begin{abstract}
In this paper, we consider hard-wall AdS/QCD models extended by a string-theory inspired Chern-Simons action in terms of a superconnection involving a bi-fundamental scalar field which corresponds to the open-string tachyon of brane-antibrane configurations and which is naturally identified with the holographic dual
of the quark condensate in chiral symmetry breaking.
This realizes both the axial and chiral anomalies of QCD with a Witten-Veneziano
mechanism for the $\eta'$ mass in addition to current quark masses,
but somewhat differently than in the Katz-Schwartz AdS/QCD model used
previously by us to evaluate pseudoscalar and axial vector transition form factors
and their contribution to the HLBL piece of the muon $g-2$.
Compared to the Katz-Schwartz model, we obtain a significantly
more realistic description
of axial-vector mesons with regard to $f_1$-$f_1'$ mixing and
equivalent photon rates. Moreover, predictions of the $f_1\to e^+e^-$ branching
ratios are found to be in line with a recent phenomenological study.
However, pseudoscalar transition form factors compare less well
with experiment; in particular the $\pi^0$ transition form factor turns out to be 
overestimated
at moderate non-zero virtuality.
For the combined HLBL contribution to the muon $g-2$
from the towers of axial vector mesons and excited pseudoscalars
we obtain, however, a result very close to that of the Katz-Schwartz model.
\end{abstract}
%\pacs{11.25.Tq,13.25.Jx,14.40.Be,14.40.Rt}

\maketitle

\section{Introduction}

The anomalous magnetic moment of the muon has been measured to date with a
world-average error of 0.19 ppm \cite{Muong-2:2023cdq,Muong-2:2024hpx}, corresponding to a standard deviation of $22\times 10^{-11}$ in $a_\upmu=(g-2)_\upmu/2$, and this uncertainty is expected to be further reduced
by about a factor of 2 with the final result of the current experiment
at Fermilab by the Muon $g-2$ Collaboration. % .
The Standard Model prediction of $a_\upmu$ is limited by uncertainties
in the contribution from hadronic vacuum polarization (HVP),
where data-driven and lattice calculations aim at per-mille accuracy to match
the experimental precision, but currently deviate from each other
beyond their respective error estimates \cite{Colangelo:2022jxc}.
Once this discrepancy is resolved,
the uncertainty of the
much smaller hadronic light-by-light (HLBL) scattering contribution
will be again of comparable importance as it contributes an error of $19\times 10^{-11}$
according to the 2020 White Paper of the Muon $g-2$ Theory Initiative \cite{Aoyama:2020ynm}.
This uncertainty is dominated by the effect of
short-distance constraints (SDCs) 
\cite{Melnikov:2003xd,Bijnens:2019ghy,Bijnens:2020xnl,Colangelo:2019lpu,Colangelo:2019uex,Ludtke:2020moa,Colangelo:2021nkr,Knecht:2020xyr,Bijnens:2021jqo,Bijnens:2022itw}
and axial vector contributions, for which
errors of $10\times 10^{-11}$ and $6\times 10^{-11}$, respectively, have been
estimated in \cite{Aoyama:2020ynm},
\1
corresponding to 67 and 100 \% of the assumed values.\footnote{\1The 100\% uncertainty in the axial-vector contribution is due to the
fact that conventional hadronic models \cite{Melnikov:2003xd,Pauk:2014rta,Jegerlehner:2017gek,Dorokhov:2019tjc,Masjuan:2020jsf,Radzhabov:2023odj} disagree even in the order of magnitude,
with some of them yielding contributions above $10^{-10}$, some less than $10^{-11}$.}
\0

In hadronic models, SDCs can typically only be satisfied by infinite towers
of resonances. The estimate of their effects in \cite{Aoyama:2020ynm}
was based on a Regge model of excited pseudoscalars subjected to
available experimental constraints \cite{Colangelo:2019uex,Colangelo:2019lpu}, although
in the chiral limit excited pseudoscalars decouple from the axial anomaly
responsible for the longitudinal Melnikov-Vainshein SDC; the alternative
of excited axial vector mesons was not considered because of
technical difficulties and sparse experimental data.
However, in \cite{Leutgeb:2019gbz,Cappiello:2019hwh} it was shown that in
chiral bottom-up holographic QCD (hQCD) models the longitudinal
SDC is naturally satisfied by the infinite tower of axial-vector mesons.
Remarkably, the short-distance behavior of their transition form factors (TFFs)
obtained in \cite{Leutgeb:2019gbz}
agrees exactly with the recently derived asymptotic behavior in a light-cone expansion
\cite{Hoferichter:2020lap}
in analogy to the Brodsky-Lepage limit of the TFF of pseudoscalars, for which
such an agreement was noted already in \cite{Grigoryan:2008up}. In the case of the individual TFFs,
it is the infinite tower of vector resonances 
of holographic models
which is responsible for this feature, while
for the complete HLBL four-point function, the infinite tower of axial vector mesons
is required for satisfying the longitudinal SDC.
While the simplest chiral hQCD models involve only Goldstone bosons and no
excited pseudoscalars, also in the slightly more complicated models
that admit massive pions, it was found that the longitudinal SDC is
saturated by the tower of axial vector mesons as opposed to the tower
of excited pseudoscalars \cite{Leutgeb:2021mpu}.

In \cite{Leutgeb:2022lqw} we have employed the model of Katz and Schwartz (KS) \cite{Katz:2007tf,Schafer:2007qy,Hong:2009zw}, where in addition to quark masses the effects of the
U(1)$_A$ anomaly are incorporated, 
%accounting for the relatively large mass of the $\eta'$ mass, 
to extend the numerical predictions of flavor-symmetric
hard-wall hQCD models to the contributions of the heavier $\eta^{(\prime)}$ mesons
(with a pseudoscalar glueball mixing in) and
$f_1^{(\prime)}$ axial vector mesons together with their radial excitations.
By including a gluon condensate as one further parameter and
accounting for gluonic corrections to the asymptotic behavior of TFFs,
we have found that the masses of all pseudo-Goldstone bosons as well
as their two-photon couplings can be reproduced at the percent level,
yielding $a_\upmu$ contributions in complete agreement with the WP2020
estimate. 
In this particular hQCD model,
the contributions from the axial-vector and excited pseudoscalar towers amounted to
$30.5\times 10^{-11}$ and $1.6\times 10^{-11}$, respectively,
somewhat reducing the previous flavor-symmetric hard-wall
result $\sim 40\times 10^{-11}$, to be compared with
the WP2020 estimate of $21(16)\times 10^{-11}$ for axial-vector plus SDC contributions.

However, while the octet-singlet mixing in the pseudoscalar
sector and the resulting photon couplings predicted by the KS model 
are very close to experimental data,
the predicted mixing of $f_1$ and $f_1'$ axial vector mesons is
rather far from experimental findings, pointing to the need of further refinements.
Adopting some parts of the construction of improved hQCD models in
\cite{Casero:2007ae,Jarvinen:2022mys}, in this paper
we consider a simple hard-wall hQCD model where
both flavor and axial anomalies are implemented by a Chern-Simons action
that follows more closely a string-theoretic setup with
brane-antibrane configurations. Besides a somewhat different implementation
of the U(1)$_A$ anomaly than in the KS model, 
this leads to a Chern-Simons action involving
the bi-fundamental scalar field corresponding to the open-string tachyon
responsible for brane-antibrane merging and thus chiral symmetry breaking.
Remarkably, this yields a significantly more realistic description
of $f_1$-$f_1'$ mixing, the equivalent photon rates, and
the slope parameters of axial-vector TFFs, but at the price
of pseudoscalar transition form factors that now overestimate experimental
data at non-zero photon virtualities.
The resulting HLBL contributions to $a_\upmu$ are found to be
smaller for the axial vector mesons and larger for the pseudoscalars,
with rather little changes in the sum total.
While not yet a completely satisfying hQCD model, we consider these results
as encouraging with regard to the 
potential of more elaborate string-inspired hQCD models
along the lines of Refs.~\cite{Casero:2007ae,Jarvinen:2022mys}
and also as providing an idea of the range of systematic errors
of current hQCD results for the HLBL contribution to the muon $g-2$.

This paper is organized as follows.
In Sec.\ 2 we give a brief introduction to the tachyon condensation models of \cite{Casero:2007ae,Jarvinen:2022mys} which are inspired by brane-antibrane constructions in string theory. 
In Sec.\ 3 we describe explicitly the HW AdS/QCD model with relevant parts of the scalar-extended Chern-Simons term included, before detailing
the equations of motion and the normalizations for its solutions for both pseudoscalar and axial vector resonances in Sec.\ 4.
In Sec.\ 5 and 6, we discuss and evaluate the resulting TFFs and
also the decay rate of axial vector mesons in electron-positron pairs, which
has been proposed in Ref.~\cite{Zanke:2021wiq} as an indirect check of
doubly virtual axial-vector TFFs.
We also discuss briefly scalar mesons, which in the simplest HW AdS/QCD models
do not participate in the HLBL scattering amplitude but for which
the tachyon-condensation models suggest a specific form of photon interactions.
Finally, we present numerical results for the various HLBL contributions to $a_\upmu$,
comparing the different $N_f=2+1$ hQCD models with U(1)$_A$ anomaly.

In the appendices, we describe an important matching factor and the explicit forms appearing in the Chern-Simons term in detail, and we provide details on scalar TFFs.

\section{Brane anti-brane systems and superconnections}

In this section we briefly review some features of $Dp$-$\overline{Dp}$ systems in the context of holography, closely following \cite{Casero:2007ae}. We describe some features of the resulting open-string tachyon condensation models and focus in particular on the Chern-Simons term, which we will subsequently include in a simple bottom-up hard-wall (HW) AdS/QCD model.

The string-theoretic construction is based on a configuration of $N_c$ $Dq$ branes intersecting with $N_f$ $Dp$ and $N_f$ $\overline{Dp}$ branes in type IIA or type IIB string theory. In the large $N_c$ ('t Hooft) limit, the color branes are described by background values of the massless string fields (including the graviton), while the flavor branes (which all sit on top of each other) are treated as non-backreacting probe branes. This yields a dual gauge theory with gauge group $SU(N_c)$ with matter in the fundamental representation and a global symmetry group of $U(N_f)_L \times U(N_f)_R$, whose $U(1)_A$ part will be broken by an anomaly.
The fields localized to the flavor branes will be a $U(N_f)_L \times U(N_f)_R$ gauge field $(A_L,A_R)$ and a complex scalar field $T$ in the bi-fundamental representation of the gauge group.

The low-energy effective action contains a tachyonic mass term for $T$ associated with the annihilation of $Dp$ and $\overline{Dp}$ branes into lower dimensional strings and branes, which (at least in flat space) is given by $m_T^2=-1/(2\alpha')$, with $\alpha'$ being the fundamental string tension. In order to interpret the scalar as the dual field of the quark condensate operator $\overline{\psi_L}_{\bar{i}} \psi_{R j}$, its mass has to be set to $m_T^2=-3/R^2$; hence, the AdS radius $R$ has to satisfy $R^2=6 \alpha'$. This of course signals a breakdown of the supergravity approximation,
which will nevertheless be taken as
the basis of constructing a phenomenological bottom-up model
for QCD.

Holographic QCD models from tachyon condensation \cite{Iatrakis:2010zf,Iatrakis:2010jb}(and their extensions to the Veneziano limit \cite{Jarvinen:2011qe,Jarvinen:2022mys}) are obtained by prescribing an AdS-like background, setting $R^2=6 \alpha'$ and ignoring all higher string theory modes (or including their effects in various different effective potentials). They are semiclassically self-consistent, and the scalar $T$ obtains a stable background value (i.e.\ condenses), which describes chiral symmetry breaking in the boundary theory.

These models contain a particularly interesting generalization of the usual Chern-Simons term that involves a superconnection.\footnote{This generalization has been recently shown to be the appropriate framework for a wider class of anomalies including also systems with interfaces and spacetime boundaries \cite{Cordova:2019jnf,Kanno:2021bze}.} The world-volume action of the flavor and anti-flavor branes splits into two parts, $S=S_{DBI}+S_{CS}$, where the Chern-Simons part is given by %(note that the scalar $T$ here has mass dimension 1)
    \begin{align}
    \label{eq:CSterm}
    S_{CS}&=T_p \int_{\Sigma_{p+1}}C \wedge \text{Str} \exp[{i2\pi \alpha' \mathcal{F}} ]    
\end{align}
with
\begin{align}
     i\mathcal{F}&=\begin{pmatrix}
iF_L-T^{\dagger}T & DT^{\dagger} \\
DT & iF_R -TT^{\dagger} \end{pmatrix},
 \\
C&= \sum_n (-i)^{\frac{p-n+1}{2}} C_n \nonumber.
\end{align}
Here $C$ is a formal sum of Ramond-Ramond (RR) $n$-form potentials $C_n$
and the integral is supposed to pick the $(p+1)$-form part of the integrand.

Contrary to the DBI action, the Chern-Simons action can be derived from boundary string field theory to all orders in $\alpha'$ \cite{Kraus:2000nj,Takayanagi:2000rz}.
The above structure actually has a very geometric origin in Quillen's theory of superconnections on  $\mathbb{Z}_2$ graded vector bundles \cite{Quillen:1985vya}. $ \mathcal{F}$ is known as the curvature of the superconnection $\mathcal{A}$ and the multiplication implicit in the definition of the exponential is a $\mathbb{Z}_2$ generalization of the usual matrix multiplication. A useful aspect of Quillen's results is that when the bundle is trivial (which we will assume in the rest of the paper) $\text{Str} \exp{[i2\pi \alpha' \mathcal{F}]}=i d \Omega$ is a total derivative.
In the setup of \cite{Casero:2007ae}, the color branes generate an RR flux proportional to $N_c$, which gets picked up by the 5 form part $\Omega_5$. After effectively reducing dimensionally to $AdS_5$, there is a quadratic coupling of the $C_3$ RR potential and the fields contained in $\Omega_1$; this is responsible for the correct implementation of the $U(1)_A$ anomaly. 
Hence, the two most important terms from the Chern-Simons action will be
\begin{equation}\label{eq:twoCSs}
    T_p \int_{AdS_5} C_{3} d \Omega_1+\frac{ N_c}{(2\pi)^2}\frac{1}{(2 \pi \alpha')^3} \int_{AdS_5} \Omega_{5}.
\end{equation}
In the following section we implement this structure into a hard-wall (HW) hQCD model.
\1
While the first term is subleading in $N_c$, it plays an important role phenomenologically, as it will be responsible for the large mass of the $\eta'$ meson.
\0

\section{Hard-wall AdS/QCD %inspired bottom up 
model with %new 
scalar-extended Chern-Simons term}

Upon expanding the DBI action of the tachyon condensation models, one obtains in lowest order of $\alpha'$ exactly the action of the original HW AdS/QCD models of Ref.~\cite{Erlich:2005qh,DaRold:2005mxj} which are known to be well suited for the description of low lying mesons.
The tachyon $T$ can be identified with the bi-fundamental scalar $X$ of the HW models. There is, however, an important factor that determines the interactions in the Chern-Simons term involving the scalar fields.
The relation, which is derived in Appendix \ref{app:TX}, is 
\begin{equation}
\label{eq:confact}
    T=X^{\dagger} g_5 \sqrt{\frac{\pi}{2}},
\end{equation}
where $g_5$ is the 5-dimensional Yang-Mills coupling of flavor gauge fields
and $X$ is a bi-fundamental scalar field dual to bilinear quark operators.
Since the tachyon has dimension 1 and $g_5$ has mass dimension $-\frac{1}{2}$, we obtain the correct mass dimension for $X$, namely $\frac{3}{2}$.

In bottom-up HW models, 
the background geometry is simply pure $AdS_5$, which (upon setting the AdS radius equal to 1) has a metric
\begin{equation}
    ds^2=z^{-2}(\eta_{\mu\nu}dx^\mu dx^\nu - dz^2), 
\end{equation}
in Poincaré coordinates.
Confinement and conformal symmetry breaking 
is implemented by a sharp cutoff at $z=z_0$, where one has to specify boundary conditions for the fields.

The complete action relevant to the processes that we will consider later reads
\begin{align}
\label{eq:totact}
      S&=  -\frac{1}{4g_5^2}\!\!\int\limits_{AdS_5} \!\!\!\sqrt{ g } \,\text{tr} \left[(F_L)_{MN}(F_L)^{MN}+ (F_R)_{MN}(F_R)^{MN}\right] \nonumber \\&+\int\limits_{AdS_5} \!\!\! \sqrt{g } \,\text{tr} \left[D_M X D_N X^{\dagger}g^{M N}+ 3X X^{\dagger}\right] \nonumber \\ &+\frac{ N_c}{(2\pi)^2}\frac{1}{(2 \pi \alpha')^3} \int\limits_{AdS_5} \Omega_{5} \nonumber \\ &+ \int\limits_{AdS_5} \frac{1}{2 B(z)}  d C_3 * dC_3 + T_p \int\limits_{AdS_5} C_{3}\, d \Omega_1,
\end{align} 
where it is understood that we replace $\alpha'$ by $\frac{1}{6}$ and any occurrence of $T$ in the Chern-Simons terms has to be replaced by \eqref{eq:confact}. 
The first three integrals are identical to the usual HW action upon setting $T=0$ in $\Omega_5$.

In the AdS geometry, the scalar field $X$ admits a background
of the form
\begin{equation}
    X_0=\frac{1}{2}(m z+ \sigma z^3)= \frac{1}{2}\begin{pmatrix}
        v_q & & \\
         & v_q & \\
          & & v_s
    \end{pmatrix},
\end{equation}
where the matrices $m, \sigma$ are proportional to the quark masses and the chiral condensate in the dual theory. We will restrict to $N_f=3$, with $m_u=m_d\neq m_s$ but uniform $\sigma\propto\mathbf{1}$.
These parameters together with $z_0,g_5$ will be fitted to $\pi_0,\rho_0$, $K^0$ and OPE data below.

The fluctuations around the vacuum of the scalar and the longitudinal gauge fields describe scalar and pseudoscalar mesons, while the transverse degrees of freedom of the gauge fields describe vector and axial-vector mesons.
We parametrize the pseudoscalar fluctuations of $X$ as \cite{Abidin:2009aj}
\begin{equation}
\label{eq:scalfluc}
    X= e^{i\eta}X_0 e^{i\eta},\quad \eta=\eta^a(x,z)t^a,\quad a=0,\ldots,8
\end{equation}
where $\text{tr}(t^a t^b)=\delta^{ab}/2$.

In %line 4 of 
\eqref{eq:totact} we have included a phenomenological background field $B(z)$ in the kinetic term of the RR 3-form, which will eventually account for the running of the QCD coupling in the gluon condensate. Some explicit expressions for $\Omega_1,\Omega_5$ relevant for our discussion are given in a separate appendix.
It is possible to dualize the last line of \eqref{eq:totact} to
\begin{align}
   S_{a}= 
   \int \frac{1}{2} B(z) (da-T_p \Omega_1)* (da-T_p \Omega_1),
\end{align}
where $a$ can be interpreted as a pure pseudoscalar glueball field 
(before its eventual mixing with pseudoscalar mesons).
Note that $\Omega_1$ is only defined up to an extra closed term with integer periods $\Omega_1 \rightarrow \Omega_1+ \xi$. These are exactly the gauge symmetries of a circle-valued scalar, so this action is well-defined.
As is shown in the appendix, to leading order in pseudoscalar fluctuations,
\begin{align}
  \Omega_1= 2\pi \alpha' \text{tr}\Bigl(&2 \eta \,d \exp(-2 \pi \alpha' T_0^2) \nonumber\\
  &+(A_L-A_R)\exp(-2 \pi \alpha' T_0^2) \Bigr).
\end{align}
Upon rescaling\footnote{Since $a$ is a circle-valued scalar, this rescaling would affect the normalization of its periods. We are however not interested in situations with nonzero winding.} $a$ by a factor of $4 \pi \alpha'T_p$, one arrives at
\begin{align}
    S_a&=\frac{1}{2}\tilde{B}(z)(da -\tilde{\Omega}_1)\wedge* (da-\tilde{\Omega}_1) 
\end{align}
with 
\begin{align}
        \tilde{\Omega}_1&= \text{tr}\bigg( \eta \,d \exp(-2 \pi \alpha'T_0^2)+A\exp(-2 \pi \alpha'T_0^2) \bigg),\\
    A&=\frac{A_L-A_R}{2}.
\end{align}

The action in terms of the components of the field can be written as
\begin{align}
\frac{\tilde{B}(z)}{2}\sqrt{g} g^{MN}(\partial_M a- \text{tr}(V_t A_M+\eta \partial_M V_t))\nonumber\\
\times(\partial_N a- \text{tr}(V_t A_N+\eta \partial_N V_t))
\end{align}
with $V_t=\exp(-2 \pi \alpha' T_0^2)$.
Upon expanding this action at small $z$ we find
\begin{align}
    \mathcal{L}_a&=\frac{(\tilde{B}(z)\frac{N_f}{2})}{2}\sqrt{g} g^{MN}(\partial_M \tilde{a}-A^0_M)(\partial_N \tilde{a}-A^0_N)\nonumber\\
    &\mbox{with}\quad\tilde{a}=a \sqrt{{2}/{N_f}}.
\end{align}
This is, asymptotically, %precisely - 
of the same form as the action of the KS model \cite{Katz:2007tf}, with $\sqrt{\tilde{B}(z)\frac{N_f}{2}}$ identified with 
\be
\tilde{Y}_0=\frac{C_0}{-\ln z \Lambda}-\Xi_0 z^4 \big( (\ln z\Lambda) -\frac{1  }{2}+\frac{1}{8\ln z \Lambda } \big)
\ee
of \cite{Leutgeb:2022lqw}. As explained in that paper, a fit to the OPE of QCD forces the leading part of $\tilde{Y}_0$ to be proportional to the running coupling $\alpha_s$. The subleading part is a consequence of the demanding consistency of the EOM of \cite{Leutgeb:2022lqw}; it allows us to include a non-zero gluon condensate,
parametrized by $\Xi_0$, and we will take over this explicit form to the present model.

\section{Equations of motion and meson modes}

We first list the quadratic part of the action and the equations of motion for the pseudoscalar sector that follow from it.
The pseudoscalar action for the $a=8,0$ subsector reads
\begin{align}
\label{lagr}
   \mathcal{L}_2\supset & \frac{-1}{4g_5^2 z}(\partial_M A_N^a-\partial_N A_M^a)^2
   \nonumber \\ 
    & +\frac{M_{ab}^2}{2 z^3}(\partial_M \eta^a -A_M^a)(\partial_M \eta^b -A_M^b) \nonumber \\ 
   & +\frac{\tilde{B}}{2 z^3}\bigg(\partial_M a-\text{tr}\{A_M V_t+\eta\partial_M V_t\}\bigg)^2,
\end{align}
where the spacetime indices are contracted with the 5d Minkowski metric and
\begin{equation}
    M^2_{ab} = \frac{1}{3}
    \begin{pmatrix}
        2 v_q^2+ v_s^2 & \sqrt{2}( v_q^2-v_s^2)\\
         \sqrt{2}(v_q^2-v_s^2) & v_q^2+ 2 v_s^2
    \end{pmatrix}.
\end{equation}
Flavor traces of the tachyon potential $V_t=\exp(- \pi^2 g_5^2 \alpha' X_0^2 )$
will be written as 
\begin{equation}
    \text{tr}( t^a V_t)=\m^a.
\end{equation}
In the following we will work in the $A_z=0$ gauge. The pseudoscalar fluctuations of the gauge field are parametrized as $A_{\mu}^a= \partial_{\mu} \varphi^a$.
The equations of motion in the pseudoscalar sector for $a=0,8$ read
\begin{align}
    \label{eq:EOMps}
   & \partial_z(\frac{1}{z} \partial_z \varphi^a ) + g_5^2 \frac{M_{ab}^2}{z^3} (\eta^b-\varphi^b)  +g_5^2 \frac{\tilde{B}}{z^3} (a-\varphi^b \m^b) \m^a=0,\nonumber\\
       & \partial_z\bigg[\frac{\tilde{B}}{z^3} (\partial_z a -\eta^a\partial_z \m^a  ) \bigg]  + q^2 \frac{\tilde{B}}{z^3} (a-\varphi^a \m^a)=0,\nonumber\\
        & \partial_z(\frac{M_{ab}^2}{z^3} \partial_z \eta^b )  + q^2 \frac{M_{ab}^2}{z^3} (\eta^b -\varphi^b)  \nonumber\\
        & \qquad\qquad\qquad +\frac{\tilde{B}}{z^3}(\partial_z a-\eta^b \partial_z\m^b)\partial_z\m^a=0,\nonumber\\
           & - \frac{q^2}{g_5^2}  \frac{1}{z} \partial_z \varphi^a+\frac{{\tilde{B}}}{z^3} (\partial_z a-\eta^b\partial_z \m^b ) \m^a      +\frac{M^2_{ab}}{z^3}\partial_z\eta^b=0 .
\end{align}
In these equations a sum over the index $b$ is understood.
The last equation is the $A_z$ equation of motion.

The $a=3$ sector is not affected by the addition of $S_a$, which implements the
U(1)$_A$ anomaly; the corresponding fluctuation equations are the same as
in \cite{Leutgeb:2021mpu}.

An important point is the choice of boundary conditions.
After varying the action, the boundary term that has to vanish at $z=0$ and $z=z_0$ is
\begin{align}
    \frac{q^2}{g_5^2}  \frac{1}{z} \partial_z \varphi^a(\delta \varphi^a)-\frac{{\tilde{B}}}{z^3} (\partial_z a-\eta^b\partial_z \m^b )\delta a&
    \nonumber\\
    - \frac{M^2_{ab}}{z^3}\partial_z\eta^b \delta \eta^a&=0,
\end{align}
which upon using the $A_z$ equation of motion reads
\begin{align}
      &\frac{q^2}{g_5^2}  \frac{1}{z} \partial_z \varphi^a(\delta \varphi^a
      -\delta \eta^a)
      \nonumber\\
      &-\frac{{\tilde{B}}}{z^3} (\partial_z a-\eta^b\partial_z \m^b )(\delta a-\m^a \delta \eta^a)=0.
\end{align}
In the rest of the paper we will be choosing so-called \cite{Leutgeb:2021mpu}
HW3 boundary conditions at $z=z_0$, i.e. $\varphi^a=\eta^a$,
supplemented by $a=\m^b\eta^b$.

In order to compute correlation functions of operators or masses and decay constants of particles in the dual gauge theory we need to solve the set of equations \eqref{eq:EOMps} subject to the above boundary conditions in the infrared. For the former, one has to supply boundary conditions in the UV which make the solution non-normalizable, while for the latter, one has to consider normalizable modes. Normalizable modes only exist for discrete values of $q^2$ which can be identified with the mass squared of a pseudoscalar meson. 
The inner product from which the norm follows reads
\begin{align}
    \langle \Phi_n,\Phi_m \rangle&= \int\! dz  \frac{1}{g_5^2}\frac{\partial_z\varphi^{a}_{n}\partial_z\varphi^{a}_{m}}{z}+\frac{M_{ab}}{z^3}(\eta_n^a- \varphi_n^a)(\eta_m^b- \varphi_m^b)\nonumber\\  &+\frac{\tilde{B} }{z^3}(a_n-\varphi^a_n\m^a)(a_m-\varphi^b_m\m^b).
\end{align}

The decay constants are defined as in \cite{Leutgeb:2022lqw} and read
\begin{align}
    f_n^a=\left.-g_5^{-2}\partial_z\varphi^{a}_n/z\right|_{z\to0},\\
     f_G^n=\left.\tilde{B} \partial_z a_n / z^3\right|_{z\to0}.
\end{align}

The equations of motion for the $a=0,8$ axial vector mesons read
\begin{align}
    -\partial_z(\frac{1}{z}\partial_z \xi^a) - \frac{q^2}{ z}\xi^a +g_5^2\frac{M_{ab}^2}{z^3}\xi^b+g_5^2\frac{\tilde{B}}{z^3}\m^a  \m^b\xi^b=0.
\end{align}
Their norm is given by
\begin{align}
   1= \int dz \frac{1}{g_5^2 z}\xi^a \xi^a.
\end{align}
and the decay constants are computed by
\be
F_n^a=\left.-g_5^{-2}\partial_z\xi^{a}_{n}/z\right|_{z\to0}.
\ee
% After fitting some of the parameters in the model we will use the above equations to extract numerical results for decay constants and masses of the $a=0,8$ pseudoscalar mesons and axial vector mesons.

\subsection{Parameter settings}\label{sec:parms}

The HW AdS/QCD model with scalar-extended Chern-Simons terms thus constructed
will be referred to as CS'' in the following. We shall also consider
a variant CS', where the tachyon $T$ is only appearing in $\Omega_1$, which
implements the U(1)$_A$ anomaly, but
where $\Omega_5$ involves only flavor gauge fields.\footnote{In the
$a=3$ sector, the CS' model is in fact identical to the HW3 model
considered in \cite{Leutgeb:2021mpu}.} 
These two models will be compared with the version of the KS model evaluated
by us in \cite{Leutgeb:2022lqw}, to which we refer
for detailed tables of numerical results. All three models,
for which we take $m_u=m_d\not=m_s$ and $\sigma\propto\mathbf1$,
have the same number of free
parameters, which will be fixed by $f_\pi=92.21$ MeV, $m_\rho=775.556$ MeV,
\bea
m_K^2&=&\frac12(m_{K_\pm}^2+m_{K_0}^2)-
\frac12(m_{\pi_\pm}^2-m_{\pi_0}^2)\nonumber\\
&&= (495.007 \, \mathrm{MeV})^2
\eea
and a least-square fit of $m_\eta$ and $m_{\eta'}$.

The coupling constant $g_5$ is usually fixed by the OPE of the
vector current correlator as
\be\label{g5LO}
g_5^2=12\pi^2/N_c=(2\pi)^2\quad \text{(OPE fit)},
\ee
but we shall alternatively consider matching the
decay constant of the $\rho$ meson,
which in the HW model leads to \cite{Leutgeb:2022cvg}
\be\label{g5Frho}
g_5^2=0.894(2\pi)^2\quad \text{($F_\rho$-fit)}.
\ee
The latter in fact significantly improves the holographic
result for the hadronic vacuum polarization \cite{Leutgeb:2022cvg},
and such an $\approx10\%$ reduction of $g_5^2$
is also warranted by comparing with next-to-leading order QCD results
for the vector correlator at moderately large $Q^2$ values \cite{Shifman:1978bx,Melic:2002ij,Bijnens:2021jqo}.

\subsection{Scalar mesons}

Scalar mesons are naturally present in this model as scalar fluctuations of the field $X$. With just the terms quadratic in $X$
included in \eqref{eq:totact}, the spectrum will be flavor symmetric and there is no coupling to two photons for the $a=0,3,8$ mesons.\footnote{The
string-theoretic motivation of AdS/QCD models also suggests flavor-singlet scalar and tensor mesons in the form of dilaton and metric fluctuations dual to glueball
modes, which inevitably have two-photon couplings from vector-meson dominance.
In the Witten-Sakai-Sugimoto model, their contributions to $a_\upmu$ have been
estimated \cite{Hechenberger:2023ljn} to be below the level of $10^{-12}$.
In \cite{Colangelo:2024xfh}, metric fluctuations have been used as a model
for tensor mesons in general (with adjusted coupling strength), even though
they do not naturally form flavor multiplets.}

However, if one introduces tachyon potentials also in the Yang-Mills action,
as done in the improved hQCD models of Ref.~\cite{Casero:2007ae,Gursoy:2007cb,Gursoy:2007er,Iatrakis:2010zf,Iatrakis:2010jb,Jarvinen:2011qe}, one obtains additional terms such as
\begin{align}
    - \gamma \frac{g_5^2 \pi^2}{4}\sqrt{g}\,\text{tr}\left((XX^{\dagger})^2\right) ,
\end{align}
which after expanding $X=X_0+S$ break the flavor symmetry of the excitations and introduce quark mass effects to the spectrum, and
\begin{align}
 \gamma \frac{   \pi^2 \alpha'}{4} \sqrt{g}\,\text{tr}\left(X^{\dagger}XF_L^2 +XX^{\dagger}F_R^2\right) ,
\end{align}
which leads to a nonzero TFF. We have introduced here by hand the phenomenological parameter $\gamma$ in front of both terms, whose deviation from $0$ measures how much our model differs from the original action \eqref{eq:totact};
\1
$\gamma=1$ corresponds to keeping the terms of order $\alpha'$ from \eqref{eq:Vt} in the Yang-Mills action. 
\0

The mode equations for $S_n=S_n^a t^a$ are
\begin{align}
    z^5 \partial_z \frac{1}{z^3}\partial_zS_n+{(zm_n)^2}S_n +\left({3-\frac{6g_5^5 \pi^2 \gamma X_0^2}{4}}\right)S_n=0,
\end{align}
subject to boundary conditions $S(z_0)=0$ and $S(\varepsilon) \sim \varepsilon^3$ for small $\varepsilon$.
The normalization is 
\begin{align}
    \int_0^{z_0}dz \frac{S_n^a S_n^a}{z^3}=1,
\end{align}
where the sum over $a$ is understood.

The masses that are obtained for such scalars are, however,
too large to be identified with the
scalar mesons $f_0(500)$, $a_0(980)$, and $f_0(980)$, which are of predominant interest for the HLBL contribution.
In fact, those are likely tetraquark states and
thus presumably beyond a holographic description in the large-$N$ 't Hooft limit.
%(but perhaps within improved hQCD models in the Veneziano limit \cite{Jarvinen:2011qe}).
With $\gamma=1$ and $g_5$ as in \eqref{g5LO}, the $a=3,8,0$ scalar masses
are 1.722, 1.722, and 2.084 GeV; with reduced coupling \eqref{g5Frho} slightly
smaller: 1.610, 1.610, and 2.006 GeV. 
We thus do not consider them any further here,
because with these parameters they contribute negligibly to $a_\upmu$
(only $\lesssim 10^{-12}$).
We provide, however, 
some additional pertinent comments on their TFFs in Appendix \ref{app:scTFF}.

\section{Transition form factors}

\begin{figure}
\centerline{$Q^2 F_{P\gamma^*\gamma}(Q^2,0)$ [GeV]\hfill}
\includegraphics[width=0.38\textwidth]{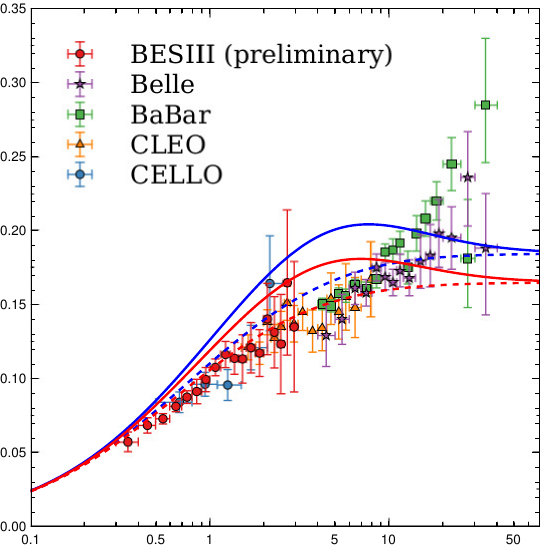}
\includegraphics[width=0.38\textwidth]{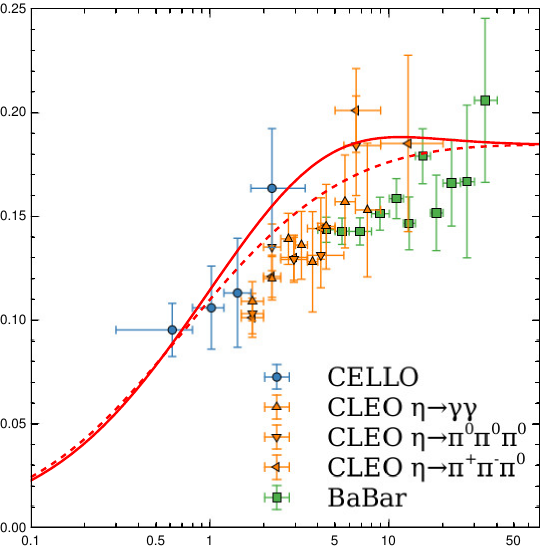}
\includegraphics[width=0.38\textwidth]{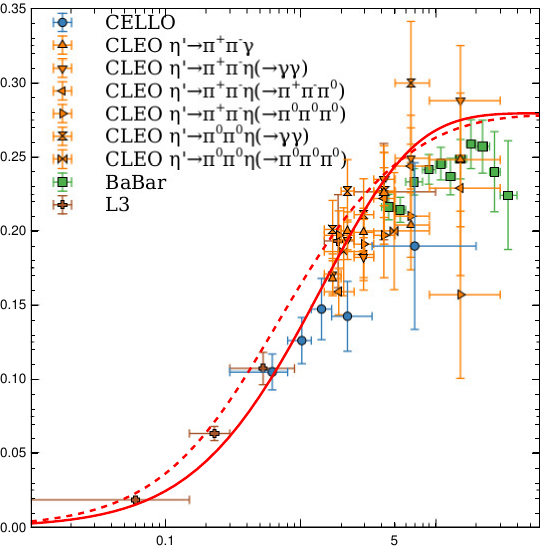}
\centerline{$Q^2$ [GeV$^2$]}
\caption{\1Singly virtual TFF $Q^2 F(Q^2,0)$ for $\pi^0,\eta$ and $\eta'$ with fully scalar extended CS term (CS'', full lines) and with scalars omitted in $\Omega_5$ (CS', dashed line), plotted on top of experimental data as compiled in Fig.~54 of Ref.~\cite{Aoyama:2020ynm}. 
Red lines correspond to reduced a reduced coupling $g_5$ ($F_\rho$ fit),
blue ones to the OPE value (only included for the experimentally
better constrained $\pi^0$ TFF).
}
\label{fig54comp}
\end{figure}

\begin{figure}
%\bigskip
\includegraphics[width=0.45\textwidth]{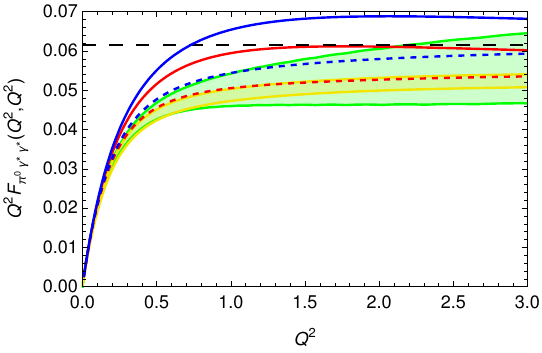}
\caption{Doubly virtual $F_{\pi^0\gamma^*\gamma^*}$ with scalar extended CS term (CS'', full line) and with scalars omitted in $\Omega_5$ (CS', dashed line), both with OPE fit (blue) and reduced $g_5$ ($F_\rho$ fit, red) in comparison with dispersive result of Ref.~\cite{Hoferichter:2018kwz} (green band) and the lattice result of Ref.~\cite{Gerardin:2019vio} (yellow band). The horizontal black dashed line indicates the BL limit.}
\label{figFpi0sconn-dvirt}
\end{figure}

Using the expressions for $\Omega_5$ derived in the appendix, we can obtain the transition form factors for the pseudoscalar mesons ($\Q$ is the quark charge matrix)
\begin{align}
    F(Q_1^2,Q_2^2)&= -\frac{N_c}{2 \pi^2} \frac{1}{3} \int dz \,\text{tr}\,\Q^2\nonumber\\
    &\times\biggl\{e^{-2 \pi \alpha' T_0^2} \left[2 \varphi'\J_1 \J_2 - \varphi(\J_1 \J_2)'\right]\nonumber\\
    &\qquad + 2  (e^{-2 \pi \alpha' T_0^2})' \varphi \J_1 \J_2 \biggr\}\nonumber\\
    &+ \frac{N_c}{2 \pi^2} \int dz \,\text{tr}\, (e^{-2 \pi \alpha' T_0^2})'\Q^2 \eta \J_1 \J_2 ,
\end{align}
where $\J_{1,2}$ is the bulk-to-boundary vector  propagator 
\be\label{HWVF}
    \J(Q_i,z)=
    Q_iz \left[ K_1(Q_iz)+\frac{K_0(Q_i z_0)}{I_0(Q_i z_0)}I_1(Q_i z) \right]
\ee
for photon virtualities $Q^2_{1,2}$ and the prime denotes differentiation with respect to $z$.
Partial integration yields
\begin{align}
\label{eq:finpstff}
    F(Q_1^2,Q_2^2)&=-\frac{N_c}{2 \pi^2} \int dz \,\text{tr}\, \Q^2 \bigg\{e^{-2 \pi \alpha' T_0^2}\varphi'\J_1 \J_2 \nonumber \\ &+(e^{-2 \pi \alpha' T_0^2})'(\varphi-\eta)\J_1\J_2\bigg\},
\end{align}
up to total derivatives which we cancel by a suitable boundary term. The quantity $\Omega_5$ is only determined up to exact terms $d\kappa$ anyway and we use that freedom to cancel the boundary term in the above equation. Doing so, the resulting TFFs satisfy a sum rule that will be derived below.

For $F(0,0)$, which determines the coupling to two real photons, we can
derive sum rules similarly to \cite{Leutgeb:2021mpu} but with certain
modifications. For those
we will need particular non-normalizable modes $\Phi_q$, which are solutions to the equations of motion for arbitrary $q^2$. Those exist only if we allow for more general boundary conditions in the UV.
The IR boundary conditions are unchanged.
We first note that the inner product of $\Phi_q$ and a normalizable mode $ \Phi_n$ reduces to a boundary contribution.
\begin{align}
    & \langle \Phi_n,\Phi_q \rangle(q^2-m_n^2)=\nonumber \\&-(\eta^a-\varphi^a)\frac{M^2_{ab}}{z^3}\partial_z\eta_n^b|_{\varepsilon}-(a-\varphi^aW^a)\frac{\tilde{B}}{z^3}(a_n'-\eta_n^bW'^b)|_{\varepsilon}.
\end{align}
For modes $\eta^0\rightarrow 1, a\rightarrow W^0(0)$ (and all other functions going to zero) the RHS becomes $m_n^2 f_n^0$.
If we let $\eta^8\rightarrow 1, a\rightarrow W^8(0)=0$ and the other functions to zero then the RHS becomes $m_n^2 f_n^8$.
We label these two different choices of boundary conditions by a lowercase (bracketed) flavor index and focus on the solution for $q^2=0$, which we denote by $\Phi_{(a)}$.
This solution will be relevant for the sum rule for the TFF and we now want to show that $\eta^a$ are constants in that case.
For $q^2=0$, the equations of motion imply
\begin{align}
    \frac{\tilde{B}}{z^3}(a'-\eta^b W'^b)=\beta,\\
    \frac{M_{ab}^2 \eta'^b}{z^3}=-\beta W^a,
\end{align}
for an as-of-yet undetermined constant $\beta$.
The solution of these equations is
\begin{align}
    \eta_{(c)}^a= \delta^a_c - \beta \int_0^z d\tilde{z}\;\tilde{z}^3(M^2_{ab})^{-1} W^b,\\
    a_{(c)}= W^c(0)+ \int_0^z d\tilde{z}\bigg( \beta \frac{\tilde{z}^3}{\tilde{B}}+\eta_{(c)}^b W'^b\bigg),
\end{align}
where $\eta$ of the first line is inserted into the second line. The first terms in each line are constants and are determined by the UV boundary conditions.
The $\varphi^a$ functions can be obtained from the above $\eta^a$ and $a$.
Using the explicit expressions in the two equations above one can see that the IR boundary condition $a(z_0)= \eta^a(z_0)W^a(z_0)$ is satisfied for $\beta=0$.
This means that $\eta^0,\eta^8$ are constant and fully determined by the boundary condition at $z=0$.

With $Z^a=\text{tr}\big( e^{-2 \pi \alpha' T_0^2} \Q^2 t^a \big) $ the TFF \eqref{eq:finpstff} at vanishing momenta reads (after using $\varphi_n^a(z_0)= \eta_n^a(z_0)$)
\begin{align}    F_n(0,0)=\frac{N_c}{2 \pi^2} \int_0^{z_0} dz Z'^a\eta_n^a 
    - \frac{N_c}{2 \pi^2}(Z^a \eta_n^a)|^{z_0},
\end{align}
hence, after multiplying by $\frac{f_n^a m_n^2}{q^2-m_n^2}|_{q^2=0}$ and summing over all modes, we can replace the normalizable mode with the non-normalizable mode.
Then the following equation is obtained
\begin{align}
    \sum_n F_n(0,0)f_n^a=\frac{N_c}{2 \pi^2} \text{tr}(\Q^2 t^a).
\end{align}

By splitting $\varphi$ into $\varphi^0 t^0 + \varphi^8 t^8$ and similar for $\eta$ we can split the TFF into two parts $F(q_1,q_2)=\bar{F}^0(q_1,q_2)+\bar{F}^8(q_1,q_2)$. To better facilitate numerical comparison with the previous analysis \cite{Leutgeb:2022lqw}, we define $F^a=\bar{F}^a/\text{tr}(t^a\mathcal{Q}^2)$ so that $F(q_1,q_2)=\sum_a \text{tr}(t^a \mathcal{Q}^2) F^a(q_1,q_2)$.

Using this decomposition we have at zero momentum 
\begin{align}    \bar{F}^a_n(0,0)=\frac{N_c}{2 \pi^2} \int_0^{z_0} dz Z'^a\eta_n^a 
    - \frac{N_c}{2 \pi^2}(Z^a \eta_n^a)|^{z_0}
\end{align}
with no sum over $a$ on the RHS.
The sum rule from before can then be generalized to (no summation over $a$ or $b$)
\begin{align}
     \sum_n \bar{F}^a_n(0,0)f_n^b=\frac{N_c}{2 \pi^2} \delta^{ab}\text{tr}(\Q^2 t^a)
\end{align}

%In order to make contact with this definition,
%we define $F^a=\tilde{F}^a/\text{tr}(t^a \Q^2)$.
%\New{(Note that the tachyon background in \eqref{eq:finpstff} has singlet and octet contributions.)}

For axial-vector mesons, we use the following parametrization\footnote{
Up to an overall factor $N_c m_A^2/(4\pi^2)$, the asymmetric
functions $A(q_1^2,q_2^2)$ corresponds to
the structure function
$\mathcal{F}_2$ of Ref.~\cite{Hoferichter:2020lap,Zanke:2021wiq},
whereas the structure function $\mathcal{F}_1$
vanishes in the hQCD models.}
for the amplitude to decay into two virtual photons
\begin{align}
\label{eq:axphph}
    \mathcal{M}^{\mu \nu \alpha}&=-i\frac{N_c}{4 \pi^2} \bigg( \varepsilon^{\tilde{\mu} \nu \rho \alpha} (q_2)_{\rho}(q_1^2\delta_{\tilde{\mu}}^{\mu}-q_1^{\mu}(q_1)_{\tilde{\mu}})A_n(Q_1^2,Q_2^2)\nonumber\\&- \varepsilon^{\mu \tilde{\nu}  \rho \alpha} (q_1)_{\rho}(q_2^2\delta_{\tilde{\nu}}^{\nu}-q_2^{\nu}(q_2)_{\tilde{\nu}})A_n(Q_2^2,Q_1^2)\bigg),
\end{align}
where $Q^2=-q^2$.
The function $A$ is determined by the Chern-Simons term and reads
\begin{align}\label{eq:A}
    A(Q_1^2,Q_2^2)=\frac{2}{Q_1^2}\int \text{tr}\Q^2 e^{-2 \pi \alpha' T_0^2} \xi \J_1' \J_2. %-(e^{-2 \pi \alpha' T_0^2}\xi \J_1 \J_2)'\bigg)
\end{align}
Below we will use the same flavor decomposition for the axial-vector TFF as for the pseudoscalar TFF.

\begin{figure}
%\bigskip
\includegraphics[width=0.46\textwidth]{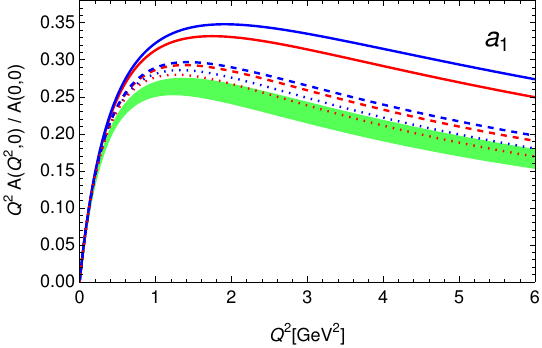}
\includegraphics[width=0.46\textwidth]{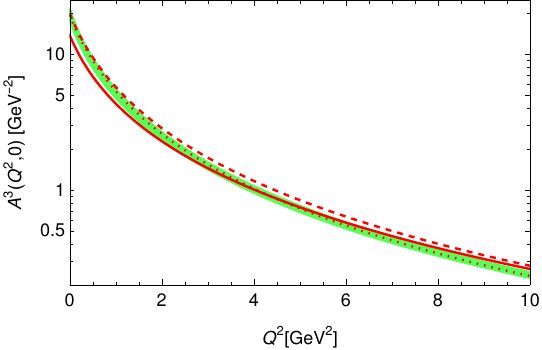}
\caption{The singly virtual $a_1$ TFF $A^3(Q^2,0)$ 
in the various models (CS'', full lines; CS' dashed lines, KS dotted lines)
and for different coupling $g_5$ (blue: OPE fit, red: $F_\rho$ fit)
in comparison to the dispersive result (green band) of Ref.~\cite{Ludtke:2024ase}.
The upper panel shows the normalized TFF $Q^2 A(Q^2,0)/A(0,0)$,
the lower panel the unnormalized $A^3(Q^2,0)$, restricted to the
reduced coupling $g_5$ obtained by the $F_\rho$ fit. Here the dotted line
representing KS($F_\rho$-fit) lies within the error band of the
dispersive result for all values of $Q^2$.}
\label{figa1singvirt}
\end{figure}

\begin{figure}
%\bigskip
%\includegraphics[width=0.45\textwidth]{a1psing-sconn3.pdf}
\includegraphics[width=0.46\textwidth]{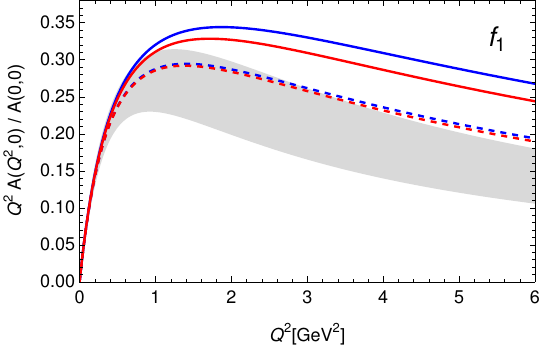}
\includegraphics[width=0.46\textwidth]{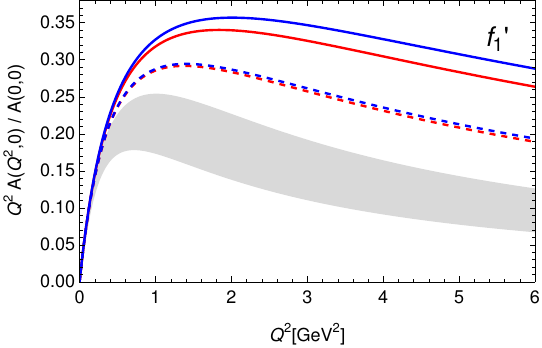}
\caption{Singly virtual TFF $Q^2 A(Q^2,0)/A(0,0)$ for $f_1$ and $f_1'$ 
(CS'': full lines, CS': dashed lines, OPE fit: blue, $F_\rho$ fit: red)
compared to L3 data \cite{Achard:2001uu,Achard:2007hm} (shaded areas). 
}
\label{f1singvirt}
\end{figure}

\subsection{Two-photon decays of axial vector mesons}

Equation \eqref{eq:axphph} together the result \eqref{eq:A}, and
$\J_1'\sim Q_1^2$ for $Q_1^2\rightarrow0$, implies that an axial vector meson
cannot decay into two real photons.
One can, however, define an equivalent two-photon rate by the
approach to zero, which in the notation of \cite{Hoferichter:2020lap,Zanke:2021wiq} reads
\be
\tilde\Gamma_{\gamma\gamma}=\frac{\pi \alpha^2 m_A}{48}|\F_\mathrm{s}(0,0)|^2,
\ee
where
\be
\mathcal{F}_\mathrm{s}(0,0)\equiv 3 m_A^2 A(0,0)/(2\pi^2).
\ee

The experimental results of \cite{Achard:2001uu,Achard:2007hm} for the singly virtual TFF of $f_1$ and $f_1'$ mesons have been
parametrized by a dipole ansatz and used in previous evaluations of
their contribution to $a_\upmu$ in \cite{Pauk:2014rta}:
\be\label{dip.}
A^\mathrm{dip.}(Q_1^2,Q_2^2)=\frac{A^\mathrm{dipole}(0,0)}{(1+Q_1^2/\Lambda_D^2)^2(1+Q_2^2/\Lambda_D^2)^2},
\ee
which can be compared with the slope parameter of the singly virtual TFF
in our calculation
\be\label{LambdaA}
\Lambda=\sqrt{\frac{-2A(Q^2,0)}{\partial A(Q^2,0)/\partial Q^2}}\bigg|_{Q^2=0}.
\ee

% The results obtained after fixing the parameters of our model as described
% in Sec.~\ref{sec:parms} are
% given in Table \ref{tab:f1experiments}.

\begin{table*}[] %%%%% pi0-a1
    \centering
\begin{tabular}{lcccccccp{10pt}ccccccc}
\toprule
& \multicolumn{6}{c}{$g_5^2=(2\pi)^2$}                                    & & \multicolumn{6}{c}{$g_5^2=0.894(2\pi)^2$}    \\ %\colrule
              & $\pi^0$ &  $\pi^*$ & $a_1$ & $a_1^*$ & {$a_1^{**}$} & $a_1^{***}$ & $a_1^{****}$  & &  $\pi^0$ &  $\pi^*$ & $a_1$ & $a_1^*$ & {$a_1^{**}$} & $a_1^{***}$ & $a_1^{****}$  \\ 
              \hline
$m$ & 0.135$^*$ & 1.709  & 1.429 & 2.419 & {3.398} &4.388&  5.3864&& 0.135$^*$ &1.579 & 1.3709 &2.366  & {3.3574} &4.356  &  5.360\\
$f\;\vee\;F_A/m_A$ & 0.09221$^*$ & 0.0016 &0.1945 & 0.244 & {0.2911} &-0.332 & 0.3695 && 0.09221$^*$ &0.0019&0.2002 &0.257 &0.3077 & 0.351 &0.390 \\ \hline
$F\vee {A^3(0,0)}$  &0.274 &0.1528 & 13.48&7.16 & {0.33}& -0.014 & 0.91 &&0.274 &0.1629 &13.86 & 6.56&-0.58 &  0.49&0.73  \\ 
%$a_\upmu\times 10^{11}$ & 73.32 & 0.9537 & 4.71 & 1.54 & 0.30 & 0.12 & 0.07 &  &68.8 & 1.09 & 4.79 & 1.42 & 0.27 & 0.13 & 0.07 \\ \hline
$a_\upmu\times 10^{11}$ & 73.3 & 0.95%37 
& 4.86 & 1.65 & 0.36 & 0.17 & 0.11 &  &68.8
 & 1.09 & 4.93 & 1.52 & 0.32 & 0.17 & 0.10 \\ \hline
$\9F\vee {\9A^3(0,0)}$ & 0.277 & 0.195 & 21.24 & -0.266 & 2.08 & 0.303 & 0.513 && 0.277 & 0.169 &20.10  &0.53 &1.80 &0.36 &0.47\\ 
$\9a_\upmu\times 10^{11}$ & 66.6 & 0.8 & 8.19 & 0.84 & 0.44 & 0.18 & 0.11 && 64.3 & 0.68 
%& 4.93 & 1.52 & 0.32 & 0.17 & 0.10 \\
& 7.59 & 0.87 & 0.40 & 0.17 & 0.10 \\
\botrule
\end{tabular}
    \caption{Results for pseudoscalar and axial-vector mesons in the isotriplet sector (the gluon condensate parameter $\Xi_0$ does not play a role here). Quantities marked by $\9{}$ correspond to the CS' model ($\Omega_5$ without extension by the scalar fields in the superconnection). All quantities in units of (powers of) GeV. }\label{tab:pia1}
\end{table*}

\begin{table*} %%%%% eta-eta'
\centering\bigskip
\begin{tabular}{lccccccp{10pt}cccccc}
\toprule
         & \multicolumn{6}{c}{$\Xi_0=0.0158$}                                 && \multicolumn{6}{c}{$\Xi_0=0.0177$}                                 \\ 
& \multicolumn{6}{c}{$g_5^2=(2\pi)^2$}                                    & & \multicolumn{6}{c}{$g_5^2=0.894(2\pi)^2$}    \\
              & $\eta$  & $\eta'$ & $G/\eta''$     & $\eta^{(3)}$ & {$\eta^{(4)}$}  & $\eta^{(5)}$  && $\eta$  & $\eta'$ & $G/\eta''$      &  $\eta^{(3)}$ & $\eta^{(4)}$ & $\eta^{(5)}$ \\ \colrule
$m$  & 0.564     & 0.944     &1.655 & 1.856    & {1.904}  &2.590  && 0.572     & 0.934   & 1.514  & 1.764     & {1.883}  & 2.521  \\
${m}-{m^\mathrm{exp}}$ &3.0\% & -1.4\% & & & & && 4.4\% & -2.5\% & & \\
$f^8$         & 0.102 & -0.034  & 0.0053 & -0.009 & {0.027} &-0.012 && 0.103 & -0.035 & 0.0055 &-0.03   & {-0.012}& 0.030\\
$f^0$         & 0.023 & 0.100 & 0.044 & {0.025} & {-0.026} &-0.0024 && 0.026 &0.105 & 0.0445 & 0.040  & {0.002} &-0.012 \\
$f_G$         & -0.029  & -0.067  & -0.129 & {-0.069}  & {0.024} &0.064 && -0.031 & -0.070  & -0.108 &  -0.059  & {0.031} &-0.053 \\ \hline
$F^8(0,0)$         & 1.48   & -0.39 &0.16  &  -0.207   & {-0.053} & 0.86&&1.47    &-0.433  & 0.166  & -0.136     & {-0.0287}&-0.968 \\
$F^0(0,0)$       &  0.45   &  1.26 &0.59 & -0.556 & {0.631} & 0.225&& 0.426  & 1.169   & 0.690  &0.128  & {0.789}&0.010   \\
$F(0,0)$           & 0.265  & 0.305 & 0.175 & -0.171   & {-0.167} &0.144 && 0.257  & 0.276  & 0.203 &  0.0218  & {0.212}& -0.09 \\
${F}-{F^\mathrm{exp}}$ & -3\% & -11\% & & & & && -6\% & -20\% & & \\
% $\theta_1$    & 66.7    & -30.9   & -72.8   & {-17.2}    & 63.8    & -33.3   & -77.7    & {-17.6}   \\
% $\theta_2$    & 61.9    & -25.4   & -88.1   & {0.12}     & 59.6    & -28.7   & -89.7    & {1.2}    \\ 
$a_\upmu\times 10^{11}$ & 19.4 & 14.9 & 1.97 & 0.87& {1.22} & 0.16 && 17.2 & 12.2  &2.6   & 0.1  & 1.62 & 0.12 \\
\hline
$\9F^8(0,0)$         & 1.57   &-0.37 &-0.68  &0.65     & {-0.77} & -0.42&& 1.56   &-0.39  & -0.64  &   0.87   & {0.127}&0.35 \\
$\9F^0(0,0)$       &  0.49   & 1.72  &-0.69 &-0.032 & {0.68} &-0.18 && 0.48  & 1.70   &  -0.67 & -0.53 & {-0.082}&  -0.13 \\
$\9F(0,0)$           & 0.284  & 0.433 & -0.253 & 0.054   & {0.11} & -0.089&& 0.282  & 0.423  & -0.245& -0.060  & -0.010& -0.0015 \\
${\9F}-{F^\mathrm{exp}}$ & +4\% & +26\% & & & & && +3\% & +23\% & & \\
% $\theta_1$    & 66.7    & -30.9   & -72.8   & {-17.2}    & 63.8    & -33.3   & -77.7    & {-17.6}   \\
% $\theta_2$    & 61.9    & -25.4   & -88.1   & {0.12}     & 59.6    & -28.7   & -89.7    & {1.2}    \\ 
$\9a_\upmu\times 10^{11}$ & 18.7 & 21.4& 0.83 &0.17 & 0.15 & 0.05 && 17.7 &  20.1& 0.89 &  0.03&0.01  &0.0003  \\
\botrule
\end{tabular}
\caption{Results for the isoscalar pseudoscalar sector, with gluon condensate values $\Xi_0$, and for two choices of $g_5$: $g_5=2\pi$ corresponding to matching the vector correlator to the LO UV-behavior in QCD, and the reduced value corresponding to a fit of $F_\rho$. The quantities $F^0,F^8$ have been defined in the text below \eqref{eq:finpstff}. Quantities marked by $\9{}$ correspond to the CS' model ($\Omega_5$ without extension by the scalar fields in the superconnection). All dimensionful quantities in units of (powers of) GeV. The central experimental value $F^\mathrm{exp}$ is 0.274 %(5) 
and 0.3437 %(55) 
GeV$^{-1}$ for $\eta$ and $\eta'$, respectively, with 3-4\% errors according to PDG \cite{ParticleDataGroup:2024cfk}. 
}
\label{tab:etas}
\end{table*}

\begin{table}[] %%%% f1-f1'
    \centering
\medskip
\begin{tabular}{lccp{12pt}cc}
\toprule
        & \multicolumn{2}{c}{$\Xi_0=0.0158$}                                 && \multicolumn{2}{c}{$\Xi_0=0.0177$}                                 \\ 
& \multicolumn{2}{c}{$g_5^2=(2\pi)^2$}                                    && \multicolumn{2}{c}{$g_5^2=0.894(2\pi)^2$}    \\
               & $f_1$ &  {$f_1'$}  && $f_1$ &  {$f_1'$}      \\ 
\colrule
$m$ & 1.503 & 1.625  &&  1.463 & 1.583 \\
%$\frac{m}{m^\mathrm{exp}}-1$ & +0.15? & +0.27? && +0.10? & +0.28? \\
$m-m^\mathrm{exp}$ & +17\% & +13\% && +14\% & +11\% \\
$m^*$ &2.450  & 2.547 && 2.402 & 2.405 \\
$m^{**}$ & 3.425 & 3.491 && 3.388 &3.450  \\ 
$F_A^8/m_A$  &0.162 &0.108 &&0.178 & 0.089\\
$F_A^0/m_A$  & 0.091&-0.139 && 0.078&-0.154 \\
\hline
$A^8(0,0)$  &13.37 &9.04 && 14.95&7.61 \\
$A^0(0,0)$  & 6.88 &-10.75&& 5.75&-11.80\\
$\theta_A$  & 62.78$^\circ$ & -40.05$^\circ$ &&68.96$^\circ$ &-32.8$^\circ$ \\
$A(0,0)$  &3.16 &-2.06&& 3.00&-2.48 \\
%$B_{e^++e^-}\times 10^{9}$ &2.51 & 0.57 && 2.02 & 0.79\\ moved to Table IV
$a_\upmu\times 10^{11}$ & 8.51 & 3.45 && 7.22 & 4.70 \\
$a^*_\upmu\times 10^{11}$ & 3.99 & 0.93 && 4.38 & 1.05 \\
$a^{**}_\upmu\times 10^{11}$ & 0.79 & 0.28 && 0.91 & 0.26 \\
$a^{***}_\upmu\times 10^{11}$ & 0.38 & 0.14 && 0.38 & 0.14\\
$a^{****}_\upmu\times 10^{11}$ & 0.26 & 0.1 && 0.24 & 0.07 \\
%$\sum a_\upmu\times 10^{11}$ \\
\hline
$\9A^8(0,0)$  &17.74 &11.63 && 17.99&8.95 \\
$\9A^0(0,0)$  & 11.69 & -17.75&&8.97 &-18.00\\
$\9\theta_A$  & 56.61$^\circ$ & -33.2$^\circ$ &&63.515$^\circ$ &-26.4$^\circ$ \\
$\9A(0,0)$  & 4.89& -3.71&& 4.17&-4.04 \\
%$\9B_{e^++e^-}\times 10^{9}$ & 4.80 & 1.60 &&3.29  &1.86 \\ moved to Table IV
$\9a_\upmu\times 10^{11}$ & 13.97 & 6.95 & & 10.42 & 8.43  \\
$\9a^*_\upmu\times 10^{11}$ & 2.14 & 0.50 && 2.61 & 0.61 \\
$\9a^{**}_\upmu\times 10^{11}$ & 0.95 & 0.30 && 1.11 & 0.27 \\
$\9a^{***}_\upmu\times 10^{11}$ & 0.41 & 0.15 && 0.39 & 0.15 \\
$\9a^{****}_\upmu\times 10^{11}$ & 0.24 & 0.1 && 0.23 & 0.07 \\
%$\sum \9a_\upmu\times 10^{11}$ \\
\botrule
\end{tabular}
    \caption{Results for the isoscalar axial-vector sector, with gluon condensate
    parameter $\Xi_0$, for the two choices $g_5$(OPE fit) and
    $g_5$($F_\rho$-fit). Quantities marked by $\9{}$ correspond to the CS' model ($\Omega_5$ without extension by the scalar fields in the superconnection);
    $\theta_A\equiv\arctan(A^8(0,0)/A^0(0,0))$
    for both $f_1$ and $f_1'$, {and $A(0,0)=\text{tr}(t^a\mathcal{Q}^2)A^a(0,0)=[A^8(0,0)+\sqrt{8}A^0(0,0)]/6\sqrt{3}$}. All dimensionful quantities are given in units of (powers of) GeV.}
    \label{tab:AV}
\end{table}

\subsection{Electron-positron decay of axial-vector mesons}

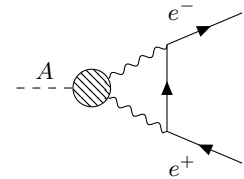
\begin{figure}[b]
    \centering
    \begin{tikzpicture}
      \begin{feynman}[small]
        \vertex (a) at (-1,0);
  \vertex[blob] (b) at (0,0) {\contour{white}{}};
\vertex (c) at (1,0.577);
\vertex (e) at (1,-0.577);
\vertex (f) at (2,1);
\vertex (g) at (2,-1);
   \diagram*{
   (a)--[scalar,edge label=$A$](b),
   (b)--[photon](c),
      (b)--[photon](e),
      (g)--[fermion,edge label=$e^+$](e)--[fermion](c)--[fermion,edge label=$e^-$](f)
   };
      \end{feynman}
    \end{tikzpicture}
   \caption{An axial vector meson decaying into electron and positron via a 1-loop process involving the doubly virtual axial-vector TFF.}    \label{fig:feynman}
\end{figure}

Axial vector mesons can also decay into an electron-positron pair
according to
an effective one-loop diagram shown in Fig.~\ref{fig:feynman}. 
The two photons emitted by the spinors couple to the axial vector meson via its TFF. This process is, therefore, a useful window into the doubly virtual axial vector TFF \cite{Zanke:2021wiq}.

Using the decomposition in terms of tensor structures of \cite{Zanke:2021wiq} the amplitude for this process can be written as
\begin{align}
\label{eq:leptdec}
    \mathcal{M}&= e^4  (\varepsilon_{\mu }\bar{u} \gamma^{\mu}\gamma^5 v) \frac{N_c}{4 \pi^2} \bigg\{ \int \frac{d^4k}{(2 \pi)^4} \frac{A_{12}+A_{21}}{k^2} \nonumber \\ &+\frac{1}{2} \int \frac{d^4k}{(2 \pi)^4} (\frac{A_{12}}{q_2^2}+\frac{A_{21}}{q_1^2})
    \bigg\}=: e^4   (\varepsilon_{\mu }\bar{u} \gamma^{\mu}\gamma^5 v) %A_1
    \bar{\mathcal{M}},
\end{align}
where $A_{12}=A(-q_1^2,-q_2^2), \; A_{21}=A(-q_2^2,-q_1^2) $ and it is understood that $q_1= p_1-k, q_2 =p_2+k$, where $p_i$ are the momenta of the fermions.
The vector $\varepsilon_{\mu}$ describes the polarization of the axial vector meson and the Dirac spinors $\bar{u}, v$ describe the spins of the electron and the positron. 
The mass of the electron which appears in $p_i^2=m_e^2$ can be neglected and we set it to zero in the following analysis.
In the decay rate, which can be compared to the experiment, one averages over the initial polarization of the axial vector meson and adds the probabilities of the different final spins of the electron and positron.

The direct numerical evaluation of the integral appearing in $\bar{\M}$ is challenging with our form for the transition form factors so we use a decomposition of the axial TFF in terms of vector meson modes
\begin{align}
\label{eq:axtffdecomp}
    A(-q_1^2,-q_2^2)=\sum_{\alpha,\beta}M_{\alpha \beta } \frac{1}{q_1^2-m_\alpha^2+i \varepsilon}\frac{1}{q_2^2-m_\beta^2+i \varepsilon},
\end{align}
with an asymmetric matrix $M_{\alpha \beta }$ defined in \eqref{eq:Malphabeta}
and
the sum running 
over the infinite tower of vector mesons with masses $m_{\alpha}$.
In the numerical evaluation (see App.~\ref{app:Mee} for more details),
it is sufficient to let $\alpha,\beta$ run over the first 60 vector meson modes, modes beyond that can be safely neglected.
The decay rate is calculated by
\begin{align}
 \Gamma_{e^+e^-}=   \frac{64 \pi^3 \alpha^4}{3}m_A |\bar{\mathcal{M}}|^2
\end{align}
and the branching ratio is obtained by dividing by the total experimental decay rate of the axial vector meson, which read
$\Gamma=23.0$ MeV for $f_1$ and $\Gamma=56.7$ MeV for $f_1'$ \cite{ParticleDataGroup:2024cfk}. The results are collected in table \ref{tab:f1experiments}.

\begin{table*}[] %%%% f1-f1' experimental data comparisons
    \centering
\medskip
\begin{tabular}{lcccp{12pt}ccc}
\toprule
        & \multicolumn{3}{c}{$\Xi_0=0.0158$}                                 && \multicolumn{3}{c}{$\Xi_0=0.0177$}                                 \\ 
& \multicolumn{3}{c}{$g_5^2=(2\pi)^2$}                                    && \multicolumn{3}{c}{$g_5^2=0.894(2\pi)^2$}    \\
              & $a_1$ & $f_1$ &  {$f_1'$}  && $a_1$ &  $f_1$ &  {$f_1'$}      \\ 
\colrule
$m^\mathrm{exp}$ & 1.23(4) & 1.2818(5) & 1.4285(14) && 1.23(4) & 1.2818(5) & 1.4285(14) \\
$m$ ($=\9m$) & 1.429 & 1.503 & 1.625  && 1.371 & 1.463 & 1.583 \\
\1 $|A(0,0)|$ (L3 exp.) & & 3.5(4) & 2.6(4) && & 3.5(4) & 2.6(4) \\
$|\mathcal{F}_\mathrm{s}(0,0)|$ (L3 exp.) & & 0.88(10) & 0.8(1) && & 0.88(10) & 0.8(1)\\
$\mathcal{F}_\mathrm{s}(0,0)$ & 0.695& 1.078[0.01+1.06] & -0.82[-0.35-0.47] &&0.658 &0.97[-0.01+0.98] & -0.94[-0.33-0.61]  \\
$\9{\mathcal{F}}_\mathrm{s}(0,0)$ &1.096 & 1.67[-0.02+1.69]& -1.49[-0.71-0.77] && 0.953  & 1.35[-0.08+1.43] & -1.53 [-0.63-0.90] \\
%$\frac{\mathcal{F}_{a_2}(q_1^2,q_2^2)}{q_1^2-q_2^2}\big|_{(0,0)}$&0.233 & 0+0.35& -0.12-0.15&& 0.377& 0+0.307& -0.11-0.19\\
$\Lambda_D$ (L3 exp.) & & 1.04(8) & 0.926(80) && & 1.04(8) & 0.926(80) \\
$\Lambda%=\sqrt{-2A(Q_1^2,Q_2^2)/\partial_{Q_1^2}A(Q_1^2,Q_2^2)}\big|_{(0,0)}
$ & 1.045 &1.043 &1.051 &&1.035 &1.033& 1.041\\
$\9\Lambda%=\sqrt{-2\9A(Q_1^2,Q_2^2)/\partial_{Q_1^2}\9A(Q_1^2,Q_2^2)}\big|_{(0,0)}
$ & 1.009 &1.007 &1.006 && 1.006 &1.005 &1.005\\
% $\tilde\Gamma_{\gamma \gamma}$ [keV] (L3 exp.) & & 3.5(8) & 3.2(9) &&  & 3.5(8) & 3.2(9) \\
% $\tilde\Gamma_{\gamma \gamma}$ [keV]&2.4 $|$ 1.4 & 6.1 $|$ 2.7 & 3.8 $|$ 2.0 &&1.9 $|$ 1.1 & 4.6 $|$ 2.5 &4.9 $|$ 2.9 \\
% $\9{\tilde\Gamma}_{\gamma \gamma}$ [keV]&2.4 $|$ 1.4 & 6.1 $|$ 2.7 & 3.8 $|$ 2.0 &&1.9 $|$ 1.1 & 4.6 $|$ 2.5 &4.9 $|$ 2.9 \\
$B_{e^+e^-}\times 10^{9}$ (SND exp.) & & $5.1_{-2.7}^{+3.7}$ &  && & $5.1_{-2.7}^{+3.7}$ & \\ 
$B_{e^+e^-}\times 10^{9}$ & &2.48 & 0.55 && & 2.0 & 0.76\\ 
$\9B_{e^+e^-}\times 10^{9}$ & & 4.74 & 1.54 && & 3.25  &1.79 \\ 
\botrule
\end{tabular}
    \caption{%Values for the quantities relevant in determining $C_s,C_{a_1}$ of \cite{Hoferichter:2023tgp}. For $f_1,f_1'$ we write the result as a sum of the $I=0$ singlet contribution and the $I=1$ contribution. For the equivalent two photon decay rate, the number after the vertical bar %$|$  is obtained by using the holographic TFF and the experimental mass, while the first number is calculated with only holographic ingredients.
    Experimental data on $f_1$ and $f_1'$ compared with the hQCD predictions in
    the CS'' model, and also the CS' model (marked by a breve sign $\9{}$ ). The equivalent photon rates measured by the L3 collaboration \cite{Achard:2001uu,Achard:2007hm} are expressed in terms of $|A(0,0)|$
    as listed in Table \ref{tab:AV} and also in terms of the parameter
    used in \cite{Hoferichter:2023tgp},
    $\mathcal{F}_\mathrm{s}(0,0)\equiv 3 m_A^2 A(0,0)/(2\pi^2)$,
    where for $f_1,f_1'$ the result is also written as a sum of the $I=0$ singlet and the $I=1$ contribution.
    The parameter $\Lambda_D$ in the dipole fit of the singly virtual TFF
    corresponds to $\Lambda$ in \eqref{LambdaA}.
    %=\sqrt{-2A(Q_1^2,Q_2^2)/\partial_{Q_1^2}A(Q_1^2,Q_2^2)}\big|_{(0,0)}$.
    The branching ratio for $f_1\to e^+e^-$ has been measured by the SND
    collaboration \cite{SND:2019rmq}.
    }
    \label{tab:f1experiments}
\end{table*}

\section{Numerical results}

Fixing the parameters of our model as described
in Sec.~\ref{sec:parms},
the numerical results for meson masses, decay constants, two-photon couplings
as well as the resulting $a_\upmu$ contributions are
listed in Table \ref{tab:pia1} for pions and $a_1$ mesons and their first few
excited modes.
In Table \ref{tab:etas} the same is done
for $\eta^{(\prime)}$ pseudoscalar mesons,
where $G/\eta''$ refers to the mode resulting from the pseudoscalar glueball
mixing with $\eta$ mesons,
and in Table \ref{tab:AV} for $f_1^{(\prime)}$ axial-vector mesons,
together with a break-up in octet and scalar contributions of decay constants
and two-photon couplings.
We give numerical results for both the fully scalar-extended CS term (CS'') 
and the partially scalar-extended one (CS'), marking results for the latter with
a breve symbol ( $\9{}$ ).
The correspondingly detailed results for the KS model are not reproduced here; for those
we refer the reader to \cite{Leutgeb:2022lqw} and list the final results for
the $a_\upmu$ contribution only in the summary table \ref{tab:total}.

\subsection{Ground-state pseudoscalars}

In the pseudoscalar sector, the achieved agreement with the experimental
data was remarkably good in the KS model (v1), 
at the few-percent level, including the two-photon couplings
$F(0,0)$.
\1
This is still the case in 
the CS'' and CS' models, except for
$F_{\eta'}(0,0)$, where larger deviations occur,
with CS'' underestimating and CS' overestimating $\eta'\to\gamma\gamma$.

The decay constants and their mixing in the $\eta$-$\eta'$ sector as
listed in Table \ref{tab:etas} require a two-angle scheme
as introduced in
\cite{Leutwyler:1997yr,Escribano:2005qq}:
\begin{align}
    \theta_8=\arctan(f^8_{\eta'}/f^8_\eta), \quad
    \theta_0=-\arctan(f^0_\eta/f^0_{\eta'}).
\end{align}
Comparing with the lattice results of
\cite{Bali:2021qem}, the results of the KS model (v1) agreed
very well with the $f^8$ scale and reasonably with $\theta_8$,
with somewhat larger deviations for the $f^0$ scale and $\theta_0$.
The CS'' and CS' models are similar, with a little larger deviation
in the octet contributions, but with a noticeable improvement in
the $f^0$ scale, which is now within the range of the (in this
case renormalization-scale
dependent) lattice result, and some improvement in $\theta_0$.
Ref.~\cite{Bali:2021qem} also determined the gluonic contribution $f_G$
(corresponding to $\sqrt{N_f/2}a$ in \cite{Bali:2021qem}), with
a scale dependent ratio $a_{\eta'}/a_\eta
\equiv f_{G,\eta'}/f_{G,\eta}$ between 2 and 2.5.
While the KS result of about 2.6 is somewhat above this range, 
the CS'' and the CS' model give 2.3, well within.
Hence, some features that are clearly associated with the U(1)$_A$ anomaly 
are more successfully reproduced in the scalar-extended CS models.
\0

\subsection{Ground-state axials}

\1
Recently, in \cite{Ludtke:2024ase}, a prediction for
the singly virtual TFF of the $a_1$ meson was obtained in the
dispersive approach, with 
\be
\mathcal{F}_\mathrm{s}^{a_1,\mathrm{disp.}}(0,0)=0.76(10).
\ee
In this case, the CS'' model is at the lower end of this prediction,
while the CS' model is 2-3 standard deviations higher.
A somewhat different picture arises if one compares the
dimensionful parameter $A^3(0,0)$, because the holographic results
for $m_A$ differs from the experimentally observed mass.
The prediction of \cite{Ludtke:2024ase} then reads
(with slightly reduced relative errors [Peter Stoffer,
private communication])
\be
A^3(0,0)^\mathrm{disp.}=19.8(1.8)\,\text{GeV}^{-2}
\ee
in terms of the values
$A^3=6A$ listed in Table \ref{tab:pia1}, which are in complete agreement in the CS' case, whereas the CS'' results are now 2-3 standard deviations too low.
With regard to $A^3_{a_1}(0,0)$, the best match is, in fact,
provided by the KS($F_\rho$-fit) model with $19.46\,\text{GeV}^{-2}$.
(Note that in the $a=3$ sector, where the U(1)$_A$ anomaly plays no role,
the CS' model is equivalent to the KS model except for slightly
different IR boundary conditions.)

In the $a=8,0$ sector, the CS models provide a somewhat better motivated
implementation of the U(1)$_A$ anomaly.
While the photon coupling $F_{\eta'}(0,0)$ happens to compare less favorably
with the experimental data in these models than the KS model, 
in the axial-vector sector,
the fully scalar-extended CS'' model turns out to give
a fairly good match
of the experimentally observed equivalent photon rates
of $f_1$ and $f_1'$ mesons.
In contrast to the KS model,
the dimensionless ratio $\tilde\Gamma_{\gamma\gamma}/m_A$, which is determined
by the dimensionless value $\mathcal{F}_\mathrm{s}(0,0)\equiv 3 m_A^2 A(0,0)/(2\pi^2)$,
%(encoded in $\mathcal{F}_\mathrm{s}(0,0)$), 
is only slightly overestimated by the CS'' result,
see Table \ref{tab:f1experiments}.
Comparing instead the dimensionful parameter $|A(0,0)|$, the CS'' result is
even below the experimental value. On the other hand, the CS' result overestimates
both.

In both, the CS'' and the CS' model, the ratio of the amplitude for $f_1$
over that for $f_1'$ is larger than 1, 
in agreement with experiment \cite{Achard:2001uu,Achard:2007hm},
while this was smaller than 1 in the KS model.
This implies that the octet-singlet mixing inferred from
experimental data is in the right ballpark.
The mixing angle, defined through
$A(0,0)$
according to 
\be
\theta_A\equiv\arctan(A^8(0,0)/A^0(0,0))
\ee
and listed
in Table \ref{tab:AV}, corresponds to
\cite{Leutgeb:2022lqw} $\theta_A^\mathrm{exp.}=56(5)^\circ$
for $f_1$.\footnote{The conventional definition
of the mixing angle $\theta_A$ in terms of the dimensionless ratio
$\tilde\Gamma_{\gamma\gamma}/m_A$ or $\mathcal{F}_\mathrm{s}$
would give $\theta_A^\mathrm{exp.}=62(5)^\circ$. The hQCD result
corresponds instead to a definition in terms of the ratio $\tilde\Gamma_{\gamma\gamma}/m_A^5$.
Also the analysis of Ref.~\cite{Roig:2019reh} suggests a dependence
of $\tilde\Gamma_{\gamma\gamma}/m_A$ on $m_A$, albeit with a different power.} 
The scalar-extended models are much closer to this range than
the KS model (v1), which gave $\theta_A\sim 80^\circ$.%
% inferred from the L3 experiment \cite{Achard:2001uu,Achard:2007hm}
% through
% \be
% \gamma_f=\frac{\tilde\Gamma_{\gamma\gamma}^{f_1}/m_{f_1}}{\tilde\Gamma_{\gamma\gamma}^{f_1'}/m_{f_1'}}
% =\left| \frac{\mathcal{F}_\mathrm{s}^{f_1}(0,0)}{\mathcal{F}_\mathrm{s}^{f_1'}(0,0)}
% \right|^2
% =1.2_{-0.5}^{+0.9} \; \text{(L3 exp.)}
% \ee
% \0
% is in the right
% ballpark, in contrast to the results of the KS model,
% \1
% where $\gamma_f$ was obtained as 0.16 and 0.10 for \eqref{g5LO} and \eqref{g5Frho},
% respectively.
% In the CS'' case, these values are instead 1.72 and 1.07.
% In the CS' case, $\mathcal{F}_\mathrm{s}(0,0)$ is too high, but the
% mixing remains realistic ($\gamma_f=1.26$ and 0.78 for \eqref{g5LO} and \eqref{g5Frho}).
\0

\subsection{Transition form factors and $a_\upmu$ contributions}

In Fig.~\ref{fig54comp}, the singly virtual TFFs of the pseudoscalar mesons
are compared with compiled data. Here we find for CS'' an excess in particular for
$\pi^0$ and $\eta$, which is associated with the fact that the BL limit
turns out to be approached from above. This feature disappears in the CS' model
(included by dashed lines). The same behavior shows up in 
the doubly virtual pion TFF (Fig.~\ref{figFpi0sconn-dvirt}).
Correspondingly the contributions to $a_\upmu$
are also excessive in the CS'' model, and no longer compatible with the WP2020
estimate as displayed in the summary table \ref{tab:total}. 

\1
In Fig.~\ref{figa1singvirt}, the shape of the normalized singly virtual TFF of
the $a_1$ meson and also the absolute magnitude
of $A^3(Q^2,0)$ as obtained in the dispersive approach
of \cite{Ludtke:2024ase} is compared to the results of the
CS'', CS', and KS models. The KS model considered by us in 
Ref.~\cite{Leutgeb:2022lqw}, which in the $a=3$ sector
is identical with the HW1m model of \cite{Leutgeb:2021mpu} and
differs from the CS' model
only with respect to boundary conditions (HW3 instead of HW1),
turns out to be closest to the dispersive result, which at large $Q^2$
is due to the fact the HW1 result has a result for the
$a_1$ decay constant that agrees best with the result for the
decay constant $F_{a_1}/m_A= 168(7)$ MeV from light-cone sum rules \cite{Yang:2007zt} used in \cite{Ludtke:2024ase} to constrain
the asymptotic behavior.
In fact, in the comparison with the unnormalized $A^3(Q^2,0)$
(lower panel of Fig.~\ref{figa1singvirt}), the KS($F_\rho$-fit) result
is completely within the narrow error band of the dispersive result, for
all values of $Q^2$.
\0

In Fig.~\ref{f1singvirt}, the shape of the singly virtual TFF of $f_1$ mesons
as obtained by a dipole fit of the L3 data \cite{Achard:2001uu,Achard:2007hm}
is compared with the CS'' and CS' results, showing a good agreement
with regard to the slope at small virtualities, but a slower decay at
larger ones. 
\1
Here the KS results are not included in the comparison,
because the $f_1/f_1'$ mixing is too far from the experimental situation.
%Here the CS' results are similar to those obtained in \cite{Leutgeb:2022lqw} in the KS model.

The doubly virtual TFFs of the axial-vector mesons also determine
the branching ratio of $e^+e^-$ decays (see Fig.~\ref{fig:feynman}).
The results obtained for the CS'' and CS' models listed in Table \ref{tab:f1experiments}
are fully compatible with the experimental results for $f_1\to e^+e^-$
obtained by the
SND collaboration in Ref.~\cite{SND:2019rmq}.
The CS'' result is also compatible with the results of the phenomenological
study of Ref.~\cite{Zanke:2021wiq,Hoferichter:2023tgp} (in particular
in the fit with better $\chi^2$ obtained by leaving out $f_1\to\phi\gamma$ data).

Considering the contribution of the ground-state axial-vector mesons to
$a_\upmu$, the CS'' model, which provides a rather good fit of
equivalent photon rates and mixing angles for $f_1$ and $f_1'$,
yields $a_\upmu^{a_1+f_1+f_1'}=16.9\times10^{-11}$, 
incidentally for both the OPE fit and the $F_\rho$-fit of $g_5$.
This is
significantly lower
than the KS($F_\rho$-fit) value of $25.0\times10^{-11}$
but still way above the WP2020 value of $6(6)\times10^{-11}$.
Since the squared masses of the ground-state axials are all overestimated
in the considered models, larger values would be obtained
by manually correcting the masses to their experimental values.
In the CS''($F_\rho$-fit), where the ground-state axials have
the best match with the L3 data, such an extrapolation would give
\be
a_{\upmu,\text{CS'' extrapol.}}^{a_1+f_1+f_1'}=20.8\,[5.99+9.15+5.66]\times10^{-11}.
%\quad\text{CS'' extrapolated}
\ee
%the CS''($F_\rho$-fit), %by around 25\%, 
% a correspondingly larger
% value should be obtained, if one corrected the masses manually.
(In \cite{Leutgeb:2019gbz}, using chiral HW models, we had obtained $a_\upmu^{a_1+f_1+f_1'}=17.4(4.0)\times10^{-11}$ by manually adjusting
rates and masses to experimental values; Ref.~\cite{Masjuan:2020jsf}
estimated this contribution as $18.38\times10^{-11}$; the recent
Dyson-Schwinger analysis of Ref.~\cite{Eichmann:2024glq}
obtained $19.4(2.1)\times10^{-11}$.)

However, the reduction of the contribution from the ground-state axial-vector
mesons in the CS'' model compared to the KS model
is accompanied by a significant increase of the
contribution from the tower\footnote{The sum over excited axials is carried out using the extrapolation described in \cite{Leutgeb:2022lqw}.
Comparing with an approximate numerical evaluation using the axial vector bulk-to-bulk propagator we estimate the possible numerical error in the complete sum
over the axial-vector towers as %+0.45 and $-0.15\times10^{-11}$ for the KS($F_\rho$-fit) model and 
+0.6 and $-0.2\times10^{-11}$.}
    %{Estimated error for infinite sum over excited axials +0.6-0.2 for OPE fit, +0.45-0.15 for $F_\rho$-fit, from assuming factor 4 or 2 instead of 2.5 in extrapolation formula. Error in pseudoscalar tower, perhaps 1/8 of first excited contribution: 0.1 for KS, 0.4 for scCS?}
of excited axial-vector mesons, see Table \ref{tab:total}.
The total contribution from axial-vector mesons is only about 10\% below
that in the KS($F_\rho$-fit)-model. Including also the contribution from
excited pseudoscalars, there is even a net increase.

As already discussed in \cite{Leutgeb:2021mpu}, the two-photon coupling of
the first excited pion is much larger than the experimental bound on
$\pi(1300)$ 
(according to \cite{Colangelo:2019uex}, $|F_{\pi(1300)\gamma\gamma}|<0.0544(71) \text{GeV}^{-1}$), 
and a similar discrepancy occurs for $\eta''$ if this is
identified with $\eta(1475)$ for which the results from the L3 experiment \cite{Achard:2007hm} imply $F(0,0)=0.041(6)$ GeV$^{-1}$. In \cite{Leutgeb:2021mpu} we
interpreted the much larger values
appearing for $\pi^*$ and $G/\eta''$ in Tables \ref{tab:pia1} and \ref{tab:etas}
as reflecting the collective effect of the much
denser spectrum of excited pseudoscalars in real QCD compared to the HW models.
Since these contributions turn out to be rather model dependent,
they should perhaps be better viewed as variable
parts of the effect of the universal
LSDC. In fact, the total sum of contributions from the towers of axials
and pseudoscalars shows remarkable stability across the models considered here.
\0

\section{Discussion and Conclusions}

\begin{table*}[t]  %%% Summary
\bigskip
\centering
\begin{tabular}{lcccccccc}
\toprule
$a_\upmu^{...}\times 10^{11}$ & KS(OPE fit) &  KS($F_\rho$-fit) & CS''(OPE fit) &  CS''($F_\rho$-fit) & CS'(OPE fit) & CS'($F_\rho$-fit) & WP2020 & Ref.~\cite{Holz:2024lom} \\
 \hline
 $\pi^0$ & 66.1 &  63.4 & 73.3 & 68.8 & 66.6 & 64.3 & 62.6$^{+3.0}_{-2.5}$ \\
 $\eta$ & 19.3 &  17.6 & 19.4 & 17.2 & 18.7 & 17.7 & 16.3(1.4) & 14.7(9)\\
 $\eta'$ & 16.9 &  14.9 & 14.9 & 12.2 & 21.4 & 20.1 & 14.5(1.9) & 13.5(7)\\
 $G/\eta''$ & 0.2 &  0.2 & 2.0 & 2.6 & 0.8 & 0.9 \\
 $\sum_{PS^*}$ & 1.6 & 1.4 & 3.5 & 3.2 & 1.2 & 0.8 \\[4pt]
 \hline
 $\pi^0+\eta+\eta'$ & 102.3 & 95.9 & 107.6 & 98.2 & 106.7 & 102.1 & 93.8(4.0) & 91.2$^{+2.9}_{-2.4}$\\
 PS poles total & 104 &  97.5 & 113 & 104 & 109 & 104 &  \\
 \hline
 $a_1$ & 7.8 &  7.1 & 4.9 & 4.9 & 8.2 & 7.6 \\
 $f_1+f_1'$  & 20.0 &  17.9 & 12.0 & 12.0 & 20.9 & 18.9 \\
 $\sum_{a_1^*}$       
 & 2.5$^\dagger$ % <- 2.2
 &  2.6$^\dagger$ % <- 2.4
 & 2.5 & 2.3 & 1.7 & 1.7 \\
 $\sum_{f_1^{(')*}} $ 
 & 4.0$^\dagger$ % <- 3.6
 &  3.5$^\dagger$ % <- 3.0
 & 7.4 
 & 7.9 & 5.3 & 5.9 \\[4pt] 
\hline
 AV+LSDC total 
 & 34.3 %\sout{33.7}
 &  31.1 %\sout{30.5}
 & 26.7 & 27.0 & 36.1 & 34.0 & \\
 AV+P*+LSDC total 
 & 36.0 %\sout{35.5}
 & 32.7 % <-32.1
 & 32.2 & 32.8 & 38.2 & 35.7 & 21(16) \\
\hline
 total & 138 & 129 % <-128
 & 140 & 131 & 145 & 138 & 115(16.5) \\
 \botrule
\end{tabular}
    \caption{Summary of the results for the different contributions to $a_\upmu$ in comparison with the White Paper \cite{Aoyama:2020ynm} values
    and with the new results on $a_\upmu^{\eta,\eta'}$ from \cite{Holz:2024lom}.
    KS refers to the Katz-Schwartz model used by us in Ref.~\cite{Leutgeb:2022lqw},
    CS'' is the fully scalar-extended case, and CS' is without scalars in $\Omega_5$;
    the two choices \eqref{g5LO} and \eqref{g5Frho} for $g_5$ are denoted
    by (OPE fit) and ($F_\rho$-fit), respectively.
    The numbers marked by a dagger have been obtained using the full bulk-to-bulk
    axial vector propagator; they
    supersede the somewhat lower estimates obtained in Ref.~\cite{Leutgeb:2022lqw} from
    extrapolating sums over the first few excited modes.
    % The sum over excited axials is carried out using the extrapolation described
    % in \cite{Leutgeb:2022lqw}.
    %{Estimated error for infinite sum over excited axials +0.6-0.2 for OPE fit, +0.45-0.15 for $F_\rho$-fit, from assuming factor 4 or 2 instead of 2.5 in extrapolation formula. Error in pseudoscalar tower, perhaps 1/8 of first excited contribution: 0.1 for KS, 0.4 for scCS?}
    }
    \label{tab:total}
\end{table*}

% In this paper, we have considered the effect of extending the Chern-Simons action in simple hard-wall AdS/QCD models by a superconnection involving the open-string tachyon which can be identified with the bi-fundamental
% scalar dual to bilinear quark operators, following the construction
% in the more involved improved hQCD models of Ref.\ \cite{Casero:2007ae,Gursoy:2007cb,Gursoy:2007er,Iatrakis:2010zf,Iatrakis:2010jb,Jarvinen:2011qe},
% and we have studied the consequences for the evaluation of TFFs and
% the HLBL tensor with a view to their contribution to the anomalous magnetic
% moment of the muon.

In Ref.~\cite{Leutgeb:2022lqw} we have considered the KS model \cite{Katz:2007tf},
where only the standard 
5-form part $\Omega_5$ (and this without scalars) is included in the
Chern-Simons action and where the U(1)$_A$ anomaly is incorporated by
a slightly simpler addition to the 5-dimensional Lagrangian.
Allowing for a gluon condensate as one further parameter, this gave
an excellent match of experimental constraints on the pseudoscalar sector,
with masses and photon couplings agreeing with data at the percent level.
Evaluating the pseudoscalar TFFs, we found good agreement
with phenomenological and lattice results when using a reduced
5-dimensional gauge coupling ($F_\rho$-fit) reflecting gluonic corrections, and
the resulting HLBL contributions to $a_\upmu$ turned out to
agree perfectly with the WP2020 estimates. 
\1
(However, there is less good
agreement with the more recent evaluation of Ref.~\cite{Holz:2024lom}
which has smaller contributions from $\eta$ and $\eta'$, see Table \ref{tab:total}.)
\0

In the best-fitting KS($F_\rho$-fit) model,
the axial-vector sector, which
is responsible for satisfying the longitudinal SDC, was found
to contribute $a_\upmu^\mathrm{AV+LSDC}=30.5\times 10^{-11}$
($32.1\times 10^{-11}$ when excited pseudoscalars are included),
slightly larger but consistent with the
WP2020 estimate of $21(16)\times 10^{-11}$, see Table \ref{tab:total}.
However, in the KS model, the axial vector mesons 
in the $a=8,0$ sector are found
to have stronger deviations from experimental data: the masses
of $f_1$ and $f_1'$ mesons are 10\% and 28\% too high, respectively,
their mixing angle deviates strongly from experimental findings,
and the equivalent photon rates are too high.
On the other hand, the combined contribution of $f_1$ and $f_1'$ to $a_\upmu$
should be fairly independent of their mixing, and the excesses in masses
and photon couplings should partially compensate in $a_\upmu$.

This points to the need for further refinements, and the improved hQCD
models developed in Ref.\ \cite{Casero:2007ae,Gursoy:2007cb,Gursoy:2007er,Iatrakis:2010zf,Iatrakis:2010jb,Jarvinen:2011qe} present several options. However,
it is certainly interesting to have alternative
hQCD models which also include flavor-symmetry breaking by
quark masses and the U(1)$_A$ anomaly, while keeping the number of
free parameters at a minimum.
This is indeed the case with the HW models involving a scalar-extended
Chern-Simons term, and we have considered these extensions
in two versions, called CS'' and CS' in Table \ref{tab:total},
where CS'' refers to the case where a scalar potential appears
in both terms in \eqref{eq:twoCSs}, while in CS'
this is the case only for the first. (Without any scalar extension,
the model is still different from the KS model, but the results
are then rather similar.)

Remarkably, both versions improve the results in the axial-vector sector
with regard to masses, mixing, and photon couplings. The masses
of the ground-state axial-vector mesons are still too high, but
only up to +17\%; the $f_1$-$f_1'$ mixing angle is close to the
experimentally observed ball-park; and the values of $\mathcal{F}_s(0,0)$
determining the equivalent photon rates are almost within
the experimental error band for the CS'' model (with either full
or reduced $g_5$). The CS' model overestimates $\mathcal{F}_s(0,0)$
similarly to the KS model.
Since the masses are too high, this should in fact tend to
underestimate the contributions of the ground-state axial-vector mesons
in the CS'' model.

However, the pseudoscalar sector, where the corresponding
parameters are nearly perfect in the KS model, shows larger
deviations in the photon coupling of $\eta'$ mesons, up to -20\%
in the CS'' model and up to +26\% in the CS' one,
see Table \ref{tab:etas}.
While the CS' model has pseudoscalar TFFs that agree similarly well
with data than the KS model, they are strongly overestimated at
nonzero virtualities in the CS'' model, approaching the BL limit
from above. In the CS'' model, the $\pi^0$ contribution is thus
much higher than the WP2020 estimate, and also the excited pseudoscalars
turn out to contribute significantly more than in the other models.

Nevertheless, in the CS'' model,
the sum of pseudoscalar and axial-vector contributions
is almost unchanged compared to the KS model.

Since the KS($F_\rho$-fit) model with reduced $g_5$ 
provides the best match in the experimentally
more constrained pseudoscalar sector, we consider it still the
preferred model for making predictions for both pseudoscalar and 
axial-vector contributions
as well as associated LSDC effects in $a_\upmu$, where
the combined contribution from $f_1$ and $f_1'$ should be
relatively insensitive to their precise octet-singlet mixing.
\1
The KS($F_\rho$-fit) model appears also validated by a comparison with the
recent dispersive results for the 
singly virtual $a_1$ TFF (see Fig.~\ref{figa1singvirt}).
\0
We therefore propose to take the range
of results added by the CS'' and CS' models as systematic
error estimates within the class of HW hQCD models which include
flavor-symmetry violating effects\footnote{In the soft-wall model
of \cite{Colangelo:2023een} that also includes the strange quark mass and the U(1)$_A$ anomaly
along the lines of \cite{Casero:2007ae}, only the pseudoscalar contributions were evaluated; the axial-vector contributions obtained in \cite{Colangelo:2024xfh}
were obtained in a flavor-symmetric model.}.
For the complete tower of axial-vector mesons whose
longitudinal parts saturate the MV-SDC, we thus
obtain for the $a_\upmu$ contributions with $F_\rho$-fitted coupling \eqref{g5Frho}
\be
a_\upmu^\mathrm{AV+LSDC}=31.1_{-4.1}^{+2.9}\times 10^{-11}. %<- 30.5_{-3.5}^{+3.5}
\ee
% \be
% a_\upmu^\mathrm{AV+LSDC}=30.5_{-3.5(3.8)}^{+3.1(5.6)}\times 10^{-11},
% \ee
% where the range before the parenthesis refers to the reduced coupling \eqref{g5Frho},
% while the one in parentheses also includes the unreduced OPE-fit \eqref{g5LO}.
[The recent estimate obtained in \cite{Eichmann:2024glq} in
a DSE/BSE calculation
of $27.5(3.2)\times 10^{-11}$ happens to be fully
compatible with this hQCD result.\footnote{Note, however, that the approximations
used therein neglect the U(1)$_A$ anomaly, which also affects the
axial-vector sector in the hQCD models.} This is also the case
for the simple Regge-like model for axial vector mesons constructed in Ref.~\cite{Masjuan:2020jsf},
which produced an estimate of $31.41\times 10^{-11}$.]

Adding the excited pseudoscalars (including $G/\eta''$), which also contribute to the
part of the HLBL tensor that is involved in the MV-SDC, the KS value
is somewhat higher and the downward
variation is reduced to zero,\footnote{\1In \cite{Leutgeb:2021mpu}, also the effect
of a larger reduction of $g_5^2$ by 15\% instead of 10\% was considered, which diminishes the
axial-vector contributions by an additional 3\%. Including that would give a
downward variation of about $1\times10^{-11}$.}
\be
a_\upmu^\mathrm{AV+P*+LSDC}=32.7_{-0.0}^{+3.0}\times 10^{-11}.
\ee
% \be
% a_\upmu^\mathrm{AV+P*+LSDC}=32.1_{-0.0(0.1)}^{+3.1(6.0)}\times 10^{-11}.
% \ee
% (Errors in parenthesis correspond to the range obtained
% by including also results with unreduced $g_5$.)
It is presumably this latter result which is to be compared with the WP2020
estimate of $21(16)\times 10^{-11}$ for the sum of axial-vector and SDC contributions,
as the estimate for the latter involved a model for excited pseudoscalars.
It should be noted, however, that the hQCD results only capture fully the LSDC;
the SDC for $\bar\Pi_1$ in the symmetric high-momentum limit has
the correct $Q^2$ behavior, but reaches only 81\% of the OPE value.\footnote{As shown
recently in Ref.~\cite{Mager:2025pvz}, this gap can be filled by tensor meson
contributions without modifying the MV-SDC.}

% Including also the ground-state pseudoscalars $\pi^0$, $\eta$, and $\eta'$,
% the total contribution from the axial sector of the six hQCD models listed
% in Table \ref{tab:total} covers the range
% \be
% a_\upmu^\mathrm{ax.sect.}=128_{-0}^{+9(17)}\times 10^{-11},
% \ee
% to be compared with the WP2020 estimate of $115(16.5)\times 10^{-11}$.

We would like to recall that the HW hQCD models we have studied are 
comparatively minimal models. They have just enough free parameters
to match $f_\pi$, $m_\rho$, and $N_f=2+1$ quark masses. The coupling $g_5$,
which is usually fixed by the leading-order OPE result for the vector correlator,
has been allowed to be replaced by a fit of $F_\rho$, which happens to
reproduce typical NLO corrections in the vector correlator as well as
in TFFs. The only extra freedom introduced was a nonzero value for the
gluon condensate that turned out to permit accurate fits of $\eta$ and $\eta'$
masses.
The three models KS, CS'', and CS' differ in their implementation
of the U(1)$_A$ anomaly and whether the bi-fundamental field is included in
the Chern-Simons terms. The simpler KS model led to the best fit in the
pseudoscalar sector, while the scalar-extended CS''
achieved the best fit to $f_1$ and $f_1'$ mesons, however at the expense
of poor agreement of the $\pi^0$ TFF with data. 
% The latter is remedied
% in the CS' model, but then part (though not all) 
% of the good match of $f_1$ and $f_1'$ disappears.
However, for $a_\upmu^\mathrm{AV+P*+LSDC}$
the CS'' model yields almost the same result as
the KS model, which we take as a validation of the 
results obtained in the latter. 

Nevertheless, it would be interesting to see how stable this result
is upon further improvements of the hQCD model.
In \cite{Colangelo:2023een}, it was found that a simple soft-wall
model with a dilaton quadratic in $z$ permits a good fit of
masses and two-photon couplings of pseudoscalars, but strongly
overestimates the $\pi^0$ TFF, which as in the CS'' model
approaches the BL limit from above.
In fact, in other applications, it was noticed before \cite{Kwee:2007dd}
that soft-wall models are phenomenologically less successful than
HW models, and this is also the case with regard to the
HVP contribution to $a_\upmu$ \cite{Leutgeb:2022cvg}.
A more promising, but also more difficult alternative
is provided by the improved hQCD models in the Veneziano limit
of Ref.\ \cite{Casero:2007ae,Gursoy:2007cb,Gursoy:2007er,Iatrakis:2010zf,Iatrakis:2010jb,Jarvinen:2011qe}, which permit the implementation of a running coupling and which indeed 
combine features of hard-wall models through an effective cutoff of the 5-dimensional
spacetime with (asymptotically) linear Regge trajectories as in soft-wall models.
The results obtained in the present study suggest that a combined
high-precision fit of low-energy data in the pseudoscalar and axial-vector sector
seems possible, which would improve further the significance of hQCD
results for the HLBL contribution to the muon anomalous magnetic moment.

After this work has been finished, a dispersive analysis of axial-vector contributions
to $a_\upmu$ has appeared in Refs.~\cite{Hoferichter:2024vbu,Hoferichter:2024bae}
which also finds significantly larger contributions from axial vector mesons
than assumed in the White Paper of 2020 \cite{Aoyama:2020ynm}. In Appendix \ref{app:E}
these results are compared with the holographic ones with regard to different
energy regions of the HLBL amplitude, and the contributions associated with
the axial sector turn out to be quite comparable, in particular with our
``best-guess'' KS($F_{\rho}$-fit) model.\footnote{Ref.~\cite{Hoferichter:2024bae} also includes contributions from tensor mesons,
where a full dispersive treatment is not yet available. Using a simple quark
model much larger contributions than previously obtained \cite{Danilkin:2016hnh}
have been found. Tensor meson contributions have now also been worked out
in HW hQCD models, which turn out to be of comparable magnitude but different sign
\cite{Cappiello:2025fyf,Mager:2025pvz}.}

\begin{acknowledgments}
We would like to thank Martin Hoferichter, Elias Kiritsis, 
Pablo Sanchez Puertas,
Peter Stoffer, and Marvin Zanke for
helpful discussions. 
We are particularly grateful to Peter Stoffer for providing
the detailed results for the $a_1$ TFF in the dispersive analysis
of Ref.~\cite{Ludtke:2024ase}.
This work has been supported by the Austrian Science Fund FWF, project no.\ 
PAT 7221623.
\end{acknowledgments}

%\newpage
\appendix
\section{Relation between $T$ and $X$}\label{app:TX}
In this appendix we derive the multiplicative factor relating the tachyon $T$ and the HW field $X^{\dagger}$.
We first list the relevant DBI action and definitions of \cite{Casero:2007ae} with $2 \pi \alpha'$ reinstated and with $T$ having mass dimension 1:
\begin{align}
    S_{DBI}&=-T_p\int d^{p+1}x e^{-\phi}\text{SymTr}\bigg(V_t(T^{\dagger}T)\sqrt{-\det B_L}   \nonumber \\ &+V_t(TT^{\dagger})\sqrt{-\det B_R}       \bigg) 
\end{align}
with
\begin{align}
  V_t(TT^{\dagger})=  e^{-2 \pi \alpha' T^{\dagger}T},\label{eq:Vt} \\
    B_{L,MN}=(g+B)_{MN}+2\pi \alpha'F^{L}_{MN}+\nonumber \\4 \pi \alpha'^2 \big( (D_MT)^{\dagger} D_NT+(D_NT)^{\dagger} D_MT\big), \\
    B_{R,MN}=(g+B)_{MN}+2\pi \alpha'F^{R}_{MN}+ \nonumber \\ 4 \pi \alpha'^2\big( D_MT(D_NT)^{\dagger} +D_NT (D_MT)^{\dagger} \big),
\end{align}
and
\begin{align}
\label{eq:covder}
    D_MT= \partial_{M}T+iTA^L_M-iA^R_MT.
\end{align}
%\textbf{This action agrees with \cite{Garousi:2004rd}, but it should be crosschecked with other papers as well}

The expression $\sqrt{-\det B_L}$ is mathematically ambiguous since $B_L$ has both spacetime and flavor indices and needs to be properly defined. The determinant should only act on the spacetime indices. We first pull out a factor of the determinant of the metric to arrive at 
\begin{equation}
    \sqrt{ g} \sqrt{\det (1+C)},
\end{equation}
with $C_L=B^M_{\;\; \, N} +2 \pi \alpha' (F_L)^M_{\;\; \, N}+4 \pi \alpha'^2 (D^MT^{\dagger} D_NT+D^NT^{\dagger} D_MT)$ and similar for $C_R$.
In the following, we will ignore the Kalb-Ramond field $B$.
We take the second factor to mean
\begin{equation}
\label{eq:pert def}
    \sqrt{\det(1+C)}=\exp{\bigg(  \frac{1}{2} \text{tr}_L \sum_{n=1} \frac{(-1)^{n+1}C^n}{n}\bigg)}.
\end{equation}
Here $C^n$ means that one should contract spacetime and flavor indices and the trace is only over the spacetime indices. We don't claim this is the correct full DBI action, we just use this prescription to expand in the lowest orders of $\alpha'$.
Performing this expansion, one arrives at\footnote{The definitions from before are in the mostly + convention, while the following formula has converted it already to mostly - as in the main text.}
\begin{align}
  &S_{DBI}=  \nonumber \\&-\frac{1}{4g_5^2}\int_{AdS_5} \sqrt{g } \text{tr} \bigg((F_L)_{MN}(F_L)^{MN}+ (F_R)_{MN}(F_R)^{MN}\bigg) \nonumber \\&+\frac{4 \alpha'}{g_5^2 2 \pi \alpha'} \int_{AdS_5}  \sqrt{g } \text{tr} (D_M T D_N T^{\dagger}g^{MN}+ \frac{1}{2 \alpha'}T T^{\dagger}).
\end{align}
The coupling constant is given in terms of the string theory parameters as $g_5^{-2}=(2 \pi \alpha')^2 T_p e^{-\phi_0} V(K)$, where $V(K)$ is the volume of possible compact directions $K$ that have been integrated out.
Upon taking ${1}/(2 \alpha')={3}/{R^2}$, with $R$ the AdS radius we can redefine the tachyon 
\begin{align}
    X= \frac{1}{g_5}\sqrt{\frac{2}{\pi}}T^{\dagger}
\end{align}
to arrive at the usual HW action for $X$.
The adjoint arises because of the definition \eqref{eq:covder} compared to our usual conventions for the covariant derivative of the scalar $X$. 

\section{Quadratic terms and radiative couplings in the Chern-Simons term}
In this appendix, we derive more explicit formulas for $\Omega_1$ and $\Omega_5$, at least when considering pseudoscalar fluctuations of the bi-fundamental scalar $T$ only. Many relevant formulas were already derived in \cite{Casero:2007ae}, but only in the special case of a tachyon background proportional to the identity. As we want to consider a differing strange quark mass, we must generalize these formulas. The exponential in \eqref{eq:CSterm} is defined by using the series expansion for the exponential, one crucial difference is that in the matrix multiplication there are extra signs depending on the form degrees of the components, we refer the reader to \cite{Casero:2007ae} for the relevant definitions.
We start with $\Omega_1$, which can be extracted from the 2-form part of $\text{Str} \exp{i2\pi \alpha' \mathcal{F}}$. We split $\mathcal{F}$ into background contributions and terms linear in fluctuations. 
In accordance with \eqref{eq:confact} and \eqref{eq:scalfluc} we parameterize fluctuations of the tachyon as $T=e^{-i \eta}T_0 e^{-i\eta}$.

The background contributions split into terms of different form-degree
\begin{align}
    i \mathcal{F}_{bg}=\begin{pmatrix}
        -T_0^2 & \\
         & -T_0^2
    \end{pmatrix}+\begin{pmatrix}
        & dT_0 \\
        dT_0 &
    \end{pmatrix}.
\end{align}
The 2 form part of $\text{Str} \exp{i2\pi \alpha' \mathcal{F}}$ splits into a term containing no $dT_0$ and a part containing one $d T_0$. After using the definitions for the multiplication of supermatrices and the supertrace these read
\begin{align}
    i 2 \pi \alpha' \text{tr} e^{-2 \pi \alpha' T_0^2} d(A_L-A_R)
    \end{align}
and \begin{align}
      i 2 \pi \alpha' \text{tr} \; d e^{-2 \pi \alpha' T_0^2} (-2 d \eta+(A_L-A_R)).
\end{align}
This immediately leads to an expression for $\Omega_1$:
\begin{align}
    \Omega_1=  4 \pi \alpha'  \text{tr}(e^{-2 \pi \alpha' T_0^2} A + \eta d e^{-2 \pi \alpha' T_0^2}).
\end{align}
As explained in the main text, any other choice of $\Omega_1$ (which differs by an exact term) can be absorbed in a redefinition of the scalar $a$.

For the computation of $\Omega_5$, we need to restrict to the 6-form part of $\text{Str} \exp{i2\pi \alpha' \mathcal{F}}$.
One way of computing $\Omega$ is using an explicit formula which reads
\begin{align}
\label{eq:transgre}
    \Omega(\mathcal{A})= 2 \pi \alpha' i \int_0^1 \text{Str} \exp (2 \pi  \alpha' i \mathcal{F}_t) \partial_t A_t,
\end{align}
where $A_t$ is a one parameter family of connections with $\mathcal{A}_1=\mathcal{A}$ and $\mathcal{A}_0$ having vanishing curvature.

Since we will only use this part of the action to compute transition functions for mesons which are described by modes that sit in the diagonal parts of the flavor matrices, we can take the fluctuations $\eta$ and the gauge fields $A_L,A_R$ to commute with the background $T_0$.
For the one-parameter family of superconnections we take
\begin{align}
\label{eq:1parfam}
    i\mathcal{A}_t=t\begin{pmatrix}
         i A_L & \\
        &  i A_R
    \end{pmatrix}
    +\begin{pmatrix}
         & d T^{\dagger}\\
        d T & 
    \end{pmatrix}.
\end{align}
The superconnection at $t=0$ does not have zero curvature in general, but it does not contain any $AVV$ or $\eta VV$ terms, which are relevant for radiative decays. Hence the curvature at $t=0$ does not contribute to the processes we consider.

The $\eta V V$ terms we get from \eqref{eq:transgre} read
\begin{align}
    i(2 \pi \alpha')^3 \text{tr}(d e^{-2 \pi \alpha'T_0^2} d \eta dV V).
\end{align}
The two $AVV$ contributions read
\begin{align}
    -i (2\pi \alpha')^3 \frac{1}{6} \text{tr} e^{-2 \pi \alpha'T_0^2}(A_L dA_L^2-A_R dA_R^2)
\end{align}
and
\begin{align}
    - i (2\pi \alpha')^3 \frac{2}{6} \text{tr} d e^{-2 \pi \alpha'T_0^2} dV (A_L-A_R)V.
\end{align}

In the above formulas we eventually replace the tachyon background with \eqref{eq:confact} (we also have $A_{L/R}=V\pm A$). We stress again that the above formulas are only valid for the meson modes sitting in the diagonal part of the flavor matrices. We note that we have performed no partial integrations to arrive at the above formulas.
One may calculate the TFFs for pseudoscalar and axial vector mesons by standard procedures from the above expressions.

In the term  $(da -\tilde{\Omega}_1)*(da -\tilde{\Omega}_1)$ one might also suspect radiative couplings (for example of the form $aVV$) when considering higher order fluctuations in $\tilde{\Omega}_1$. Using \eqref{eq:transgre}, it is easy to see that no such couplings can be produced (one has to use the 1-parameter family $\mathcal{A}_t=t\mathcal{A}$).

Characteristic for this Chern-Simons term are the $e^{-2 \pi \alpha' T_0^2}$ factors. They also provide at least a qualitative solution to a potential problem in the study of glueballs decaying into two photons. When considering such processes in models with the non-scalar extended Chern-Simons term one gets a factor of $\text{tr}(\Q^2)$ %($\Q$ being the quark charge matrix) 
from couplings of flavor-neutral fields to the flavor fields. This could be, for example, a coupling of the Kalb-Ramond field $B$, some RR field $C$, or the metric through the $\hat{A}$-genus to the flavor gauge fields. 
The value of $\text{tr}(\Q^2)$ depends rather strongly on the number of flavors of quarks one considers, while in real QCD the decay into two photons should not be sensitive to whether there exists a very heavy quark or not.  
The scalar extended CS term resolves this qualitatively since for a very heavy quark $e^{-2 \pi \alpha' T_0^2}$ goes to zero very quickly and only contributes at very small $z$, which roughly translates to very high energies.

\section{Scalar TFFs}\label{app:scTFF}

For scalars, the amplitude into two photons can be written as 
%(\textbf{no factors of $m_S$ like in HS})
\begin{align}
  \M^{\mu \nu} = T_1^{\mu \nu}\mathcal{F}_1 + T_2^{\mu \nu}\mathcal{F}_2 
\end{align}
with
\begin{align}
    &\mathcal{F}_1=- \gamma \frac{ 8  \pi^2 \alpha'}{4} \int \frac{dz}{z} \text{tr}\bigg(\Q^2 S(z) X_0(z) \bigg) \J(q_1,z) \J(q_2,z),\\
  &   \mathcal{F}_2=- \gamma \frac{ 8  \pi^2 \alpha'}{4} \int \frac{dz}{z} \text{tr}\bigg(\Q^2 S(z) X_0(z) \bigg) \frac{\J'(q_1,z)}{q_1^2} \frac{\J'(q_2,z)}{q_2^2}.
\end{align}
This agrees with the results obtained in \cite{Cappiello:2021vzi}, but there
a chiral model was considered, where $X_0(z)\sim z^3$ for $z\to0$.
Since we work with nonzero quark masses, we have $X_0(z)\sim z$ instead,
%the overall power of $z$ in the integrals differs from \cite{Cappiello:2021vzi}, hence 
we get a different asymptotic behavior, namely
\bea
&&\mathcal{F}_1 \propto %\sim ... 
\frac1{Q^4}\frac1{w^4}\left[ 
3-2w^2 + \frac3{2w}(1-w^2)\ln\frac{1-w}{1+w} \label{F1Sasym}
\right] \qquad\\
&&\mathcal{F}_2 \propto %\sim ... 
\frac1{Q^6}\frac1{w^4}\left[ 
3+ \frac1{2w}(3-w^2)\ln\frac{1-w}{1+w} 
\right],\label{F2Sasym}
\eea
whereas \cite{Cappiello:2021vzi} obtained $\mathcal{F}_1\sim Q^{-6}$ and $\mathcal{F}_2\sim Q^{-8}$,
with a different $w$-dependence.

Generalizing the pQCD calculations of Brodsky and Lepage (BL) for the pseudoscalar TFF, the authors of \cite{Hoferichter:2020lap} have obtained instead a result with the even weaker fall-off $\mathcal{F}_1\sim Q^{-2}$ and $\mathcal{F}_2\sim Q^{-4}$, where both asymmetry functions are proportional to a function $f^S(w)$
that also appears in the above result in \eqref{F1Sasym} but not in \eqref{F2Sasym}.

The OPE of the product of two electromagnetic currents in fact contains a scalar contribution
of the form
\be
\int d^4x\, e^{iqx} J^\mu(x) J^\nu(0)\sim 
\frac2{q^4} \left( g^{\mu\nu}q^2-{q^\mu q^\nu} \right) \bar\psi \Q^2 m \psi +\ldots
\ee
which is indeed consistent with the holographic result for nonvanishing quark masses.
This is not in contradiction with the BL result of \cite{Hoferichter:2020lap} and also the
slightly older study of \cite{Kroll:2016mbt}, since in
the limit of %$w=0$.
$q_1^\mu\to -q_2^\mu$ one has $T_2^{\mu\nu}\to -q^2 T_1^{\mu\nu}$
and the leading $1/q^2$ terms therein cancel
at this symmetric point.
However, in contrast to the situation for pseudoscalars and axial vector mesons,
there is a contradiction between the BL result and the holographic one away from
this point, including
the symmetric point with $q_1^\mu\to+q_2^\mu$, which is beyond
the applicability of the OPE analysis.

\1
As an aside, we would like to mention that a discrepancy between
the BL results of \cite{Hoferichter:2020lap} and holographic results
for TFFs was also found in the case of tensor mesons \cite{Colangelo:2024xfh}.
However, in this case, there is in fact a difference between the
operators involved: on the holographic side the dual operator to
the tensor mesons as introduced in \cite{Colangelo:2024xfh} is
the (flavor-singlet) full energy-momentum tensor, 
and not actually the operator underlying the analysis of Ref.~\cite{Hoferichter:2020lap}.
\0

% To compare to data we use \cite{Hoferichter:2020lap}:
% \begin{align}
%     \Gamma_{\gamma \gamma}= \frac{\pi}{4} \alpha_e^2 m_S^3 |\mathcal{F}_1(0,0)|^2
% \end{align}
% where $\alpha_e$ is the fine structure constant. 

% \Toni{Kroll vergleicht mit experimentellen Daten zum single virtual TFF von $f_0(980)$. Auch wenn das
% ausserhalb unseres Modells liegt, koennen wir das Abfallverhalten mit $Q^2$ grob vergleichen, und auch mit den chiralen Ergebnissen von Cappiello!}

% \begin{figure}
% \bigskip
% \centerline{$Q^2 \F_{1}(Q^2,0)/\F_1(0,0)$ \hfill}
% \includegraphics[width=0.42\textwidth]{F1Ssing+.pdf}
% \centerline{$Q^2$ [GeV$^2$]}
% \caption{Data on $f_0(980)$ from BELLE \cite{Belle:2015oin} compared with the two lightest scalar modes, where orange represents the degenerate $f_0$ and $a_0$ which are pure $n\bar n$, and blue is $f_0'$, which is pure $s\bar s$ and thus perhaps closest to $f_0(980)$, which in a quark-antiquark description is dominantly $s\bar s$ \cite{Kroll:2016mbt}. If the $X^4$ term is dropped, the scalar ground states are flavor symmetric and given by the green dotted line. The blue dashed line shows the result for $f_0(980)$ from the quark-model approach of Ref.\ \cite{Schuler:1997yw}. \Toni{TODO(?): Include results corresponding to the chiral HW1 model(s?) used by Cappiello, which should show a stronger decay for large $Q^2$. We could also simply use our results to evaluate $a_\upmu$ phenomenologically and not try to reproduce the parameter sets of Cappiello et al.}}
% \label{figF1S}
% \end{figure}

\section{Evaluation of the electron-positron decay amplitude of axial-vector mesons}\label{app:Mee}

In the evaluation of the electron-positron decay amplitude we use
the decomposition of the axial-vector TFF \eqref{eq:axtffdecomp} 
in terms of vector meson modes
with masses $m_\alpha$.
This decomposition can be derived by inserting the decompositions 
$\J(z,q)= \sum_{\alpha}\frac{f_{\alpha} {\rho}_{\alpha}(z)}{q^2-m_{\alpha}^2+i \varepsilon}$
and 
\begin{equation}
\label{eq:modedecomp}
    \frac{\partial_z \J(z,q)}{q^2}=\sum_{\alpha}\frac{1}{m_{\alpha}^2}\frac{f_{\alpha} {\rho}_{\alpha}'(z)}{q^2-m_{\alpha}^2+i \varepsilon}
\end{equation}
into the expression for $A(q_1^2,q_2^2)$.
Here the function $\rho_{\alpha}(z)$ is a vector meson mode obeying 
\begin{align}
    \partial_z \bigg( \frac{\partial_z \rho_{\alpha}}{z}\bigg) + \frac{m_{\alpha}^2\rho_{\alpha}}{z} =0
\end{align}
with normalization such that $\frac{1}{g_5^2}\int \frac{\rho^2}{z}=1$.
If one just straightforwardly takes the decomposition for $\J(z,q)$, and computes $\frac{\partial_z \J(z,q)}{q^2}$ one will get something different from \eqref{eq:modedecomp}. The expression that one would obtain in this way would be a sum that diverges at $q^2=0$ when truncated at any finite mode number. The expression \eqref{eq:modedecomp} is finite at $q^2=0$ even when truncating the sum at any finite mode number.
To derive \eqref{eq:modedecomp}, we note that $\lambda:= \frac{\partial_z\J(z,q)}{q^2}$ obeys the equation
\begin{align}
    (z(\frac{\lambda}{z})')'+q^2 \lambda=0.
\end{align}
To successfully decompose $\lambda$ we should look for modes of this equation.
The new modes $\tilde{\rho}$ will, of course, be proportional to $\rho'_\alpha$, but they will have a nontrivial prefactor $N_{\alpha}$ to account for their norm. Using $\tilde{\rho}_{\alpha}=N_{\alpha} \rho_{\alpha}'$, we have
\begin{align}
    1=\frac{1}{g_5^2}\int \frac{\tilde{\rho}_{\alpha}^2}{z}=N_{\alpha}^2 m_{\alpha}^2,
\end{align}
which implies $N_{\alpha}=\frac{1}{m_{\alpha}}$.
In order to calculate the inner product between the mode $\lambda$ and one of these normed modes, we compute
\begin{align}
    &(q^2-m_{\alpha}^2)\langle \tilde{\rho}_{\alpha} ,\lambda \rangle= (q^2-m_n^2) \int dz \frac{\tilde{\rho}_{\alpha}(z) \lambda(z)}{g_5^2 z} \\ &=-\frac{1}{g_5^2}\bigg(\tilde{\rho}_{\alpha} (\frac{\lambda}{z})'-\lambda (\frac{\tilde{\rho}_{\alpha}}{z})' \bigg)|_{\varepsilon}^{z_0}=-\frac{1}{g_5^2} \tilde{\rho}_{\alpha}'(\varepsilon)= \frac{f_{\alpha}}{m_{\alpha}}.
\end{align}
The last equality comes from using the small $z$ asymptotics $\lambda= -z \log z$ and $\tilde{\rho} \sim z$.
Combining all these ingredients one arrives at \eqref{eq:modedecomp}.
The expression for $M_{\alpha \beta}$ then reads
\begin{align}\label{eq:Malphabeta}
    M_{\alpha \beta}= 2 \frac{f_{\alpha}}{m_{\alpha}^2} f_{\beta}  \int dz  {\rho}_{\alpha}'(z) {\rho}_{\beta}(z)W^a \xi^a(z).
\end{align}

After insertion of \eqref{eq:axtffdecomp} into \eqref{eq:leptdec}, one can introduce Feynman parameters $u_1, u_2$ for each $\alpha, \beta$ and perform the integral over the loop variable $k$. In addition, one can integrate out $u_2$ analytically.
The result for the quantity $\bar{\M}$ in \eqref{eq:leptdec} is
\begin{align} %A_1
    \bar{\mathcal{M}}&=\frac{N_c}{4\pi^2} \frac{-i}{(4\pi)^2} \sum_{\alpha \beta} M_{\alpha \beta} \int_0^1 du_1
    \nonumber\\&\biggl\{-\frac{1}{m_{\beta}^2}\log\left[ -m_A^2u_1(1-u_1)+m_{\alpha}^2u_1-i\varepsilon\right]\nonumber \\&+\frac{1}{m_{\beta}^2}\log\left[(m_{\beta}^2-m_A^2u_1)(1-u_1)+m_{\alpha}^2u_1-i \varepsilon\right]\nonumber \\ + &\frac{2}{m_{\beta}^2-m_A^2u_1} \log \left[\frac{u_1m_{\alpha}^2+(1-u_1)(m_{\beta}^2-m_A^2u_1)-i\varepsilon}{u_1 m_{\alpha}^2-i\varepsilon}\right]\biggr\}.
\end{align}
This integral can be performed numerically to high precision, provided the
$i\varepsilon$ prescription is properly taken into account.
%, it is however important to keep the $i\varepsilon$ nonzero during the integration process. 
We have crosschecked our numerical results by performing a PV reduction using Feyncalc 9.3.1 \cite{Shtabovenko:2020gxv} and evaluating the standard integrals using LoopTools \cite{Hahn:1998yk} in Mathematica.
Using the PV definitions the integral of the last 3 lines of the previous equation can be expressed as
\begin{align}
&- 2 \, C_0\left( m_A^2, 0, 0, m_\alpha^2, m_\beta^2, 0 \right) \nonumber \\&- \frac{ B_0\left( m_A^2, m_\alpha^2, m_\beta^2 \right)  - B_0\left( m_A^2, 0, m_\beta^2 \right) }{ m_\alpha^2 } 
\end{align}
and directly evaluated using LoopTools.

\section{Comparison with Ref.~\cite{Hoferichter:2024bae}}\label{app:E}

For the sake of comparison with the recent study \cite{Hoferichter:2024bae} (HSZ)
of the subleading
contributions to $a_\upmu^\mathrm{HLBL}$ from the axial sector, Table
\ref{tab:HSZcomparison} 
shows the various contributions from the low-energy (IR) and mixed IR-UV regions
of integration, defined by the separation scale $Q_0=1.5$, for the
KS and CS'' models with the reduced $g_5^2$ obtained by fitting $F_\rho$.

This shows a remarkable agreement of the KS($F_\rho$-fit) model
with the dispersive results for axial-vector contributions 
in the IR region and only a small
deviation in the mixed region; the CS'' model agrees well with
regard to the total sum in the IR region, while the distribution
over the various contribution is not reproduced.

The mixed region contributions in the HSZ results \cite{Hoferichter:2024bae}
also include contributions other
than from the axial sector, in particular tensor meson
contributions, which have been found to be larger
than previously estimated. Tensor meson contributions to
$a_\upmu$ have most recently also been evaluated in hQCD for
the HW geometry shared by the KS and CS models considered here,
resulting in similarly large contributions, but with
opposite sign, see Ref.~\cite{Cappiello:2025fyf}.

Ref.~\cite{Hoferichter:2024bae} also lists longitudinal ($\bar\Pi_{1,2}$) and transverse ($\bar\Pi_{3-12}$)
parts of the various contributions. There the axial vector
contributions are 58-59\% longitudinal, where the various
axial vector contributions in the hQCD models have a 
54-55\% longitudinal fraction. 
Including the excited pseudoscalars, which are
purely longitudinal and which
are not present in the HSZ result, the overall
longitudinal part becomes 59\% in the KS($F_\rho$-fit) model,
equal to the amount in the dispersive result,
whereas the CS'' model has 70\%.

%\vfil

\begin{table}[h] 
\bigskip
    \centering
\begin{tabular}{llccccc}
\toprule
Region & $a_\upmu^{...}\times 10^{11}$ & KS($F_\rho$-fit) & CS''($F_\rho$-fit) & HSZ\cite{Hoferichter:2024bae} \\
\colrule
IR
& $a_1$ & 4.2 & 2.5 & 3.8(7) \\
& $f_1+f_1'$ & 8.9 & 5.9 & 8.4(1.4) \\
& $AV^*$ & 0.7 & 1.7 & \\
& $PS^*$ & 1.7 & 4.8 & \\
& eff.poles & & & 2.0 \\
& Sum & 15.4 & 14.8 & 14.2(1.6) \\
\hline
Mixed 
& $a_1$ & 2.4 & 1.95 \\
& $f_1+f_1'$ & 7.1 & 4.75 \\
& $AV^*$ & 1.9 & 1.7 \\
& $PS^*$ & $-0.04$ & 1.0 \\
& Sum & 11.4 & 8.4 & 15.9(1.0) \\
\hline
IR+Mixed & Sum & 26.8 & 25.2 & 30.0(1.9) \\
\botrule
\end{tabular}
%%%%%%%%%%%%%%%%%%%%%%%%%%%%%%%
% \begin{tabular}{llccccc}
% \toprule
% Region & & HSZ\cite{Hoferichter:2024bae} & KS($F_\rho$-fit) & CS''($F_\rho$-fit) \\
% \hline
% $Q_i<Q_0$ & $A=f_1,f_1',a_1$ & 7.2(1.4)+5.0(1.0)=12.2(2.3) & 13.1 \\
%     & $S=f_0(1370),a_0(1450)$ & $-0.7(3)$ & \\
%     & $T=f_2,a_2$ & $-2.5(3)$ & ? \\
%     & other %(effective poles or $A^*$, $PS^*$, \ldots) 
%     & 2.0 & 2 \\
% \hline
% Mixed & $A,S,T$, other & 15.9(1.0) & 12 \\
% \hline
% $Q_i>Q_0$ & & 6.2$^{+0.2}_{-0.3}$ & 6 \\
% \hline
% Sum & & 33.2(3.3) & 33 (+?) \\
% \botrule
% \end{tabular}
\caption{Low-energy (IR) region ($Q_i<Q_0=1.5$ GeV for all $i$) 
and mixed region contributions compared to the recent
evaluation of subleading contributions in the dispersive approach
of Ref.~\cite{Hoferichter:2024bae} (HSZ). In the mixed region, the HSZ
result is not exclusively from the axial sector but instead involves
OPE results in certain subregions as well as scalar and tensor contributions.
\label{tab:HSZcomparison}
}
\end{table}

\newpage

\raggedright
\bibliographystyle{JHEP}
\bibliography{hlbl}

\providecommand{\href}[2]{#2}\begingroup\raggedright\begin{thebibliography}{10}

\bibitem{Muong-2:2023cdq}
{\bf Muon g-2} Collaboration, D.~P. Aguillard et~al., {\it {Measurement of the
  Positive Muon Anomalous Magnetic Moment to 0.20~ppm}},  {\em Phys. Rev.
  Lett.} {\bf 131} (2023), no.~16 161802,
  [\href{http://arxiv.org/abs/2308.06230}{{\tt arXiv:2308.06230}}].

\bibitem{Muong-2:2024hpx}
{\bf Muon g-2} Collaboration, D.~P. Aguillard et~al., {\it {Detailed report on
  the measurement of the positive muon anomalous magnetic moment to 0.20~ppm}},
   {\em Phys. Rev. D} {\bf 110} (2024), no.~3 032009,
  [\href{http://arxiv.org/abs/2402.15410}{{\tt arXiv:2402.15410}}].

\bibitem{Colangelo:2022jxc}
G.~Colangelo et~al., {\it {Prospects for precise predictions of $a_\mu$ in the
  Standard Model}},  \href{http://arxiv.org/abs/2203.15810}{{\tt
  arXiv:2203.15810}}.

\bibitem{Aoyama:2020ynm}
T.~Aoyama et~al., {\it {The anomalous magnetic moment of the muon in the
  Standard Model}},  {\em Phys. Rept.} {\bf 887} (2020) 1--166,
  [\href{http://arxiv.org/abs/2006.04822}{{\tt arXiv:2006.04822}}].

\bibitem{Melnikov:2003xd}
K.~Melnikov and A.~Vainshtein, {\it {Hadronic light-by-light scattering
  contribution to the muon anomalous magnetic moment revisited}},  {\em Phys.
  Rev.} {\bf D70} (2004) 113006,
  [\href{http://arxiv.org/abs/hep-ph/0312226}{{\tt hep-ph/0312226}}].

\bibitem{Bijnens:2019ghy}
J.~Bijnens, N.~Hermansson-Truedsson, and A.~Rodr\'\i{}guez-S\'anchez, {\it
  {Short-distance constraints for the HLbL contribution to the muon anomalous
  magnetic moment}},  {\em Phys. Lett. B} {\bf 798} (2019) 134994,
  [\href{http://arxiv.org/abs/1908.03331}{{\tt arXiv:1908.03331}}].

\bibitem{Bijnens:2020xnl}
J.~Bijnens, N.~Hermansson-Truedsson, L.~Laub, and A.~Rodr\'\i{}guez-S\'anchez,
  {\it {Short-distance HLbL contributions to the muon anomalous magnetic moment
  beyond perturbation theory}},  {\em JHEP} {\bf 10} (2020) 203,
  [\href{http://arxiv.org/abs/2008.13487}{{\tt arXiv:2008.13487}}].

\bibitem{Colangelo:2019lpu}
G.~Colangelo, F.~Hagelstein, M.~Hoferichter, L.~Laub, and P.~Stoffer, {\it
  {Short-distance constraints on hadronic light-by-light scattering in the
  anomalous magnetic moment of the muon}},  {\em Phys.\ Rev.\ D} {\bf 101}
  (2020) 051501, [\href{http://arxiv.org/abs/1910.11881}{{\tt
  arXiv:1910.11881}}].

\bibitem{Colangelo:2019uex}
G.~Colangelo, F.~Hagelstein, M.~Hoferichter, L.~Laub, and P.~Stoffer, {\it
  {Longitudinal short-distance constraints for the hadronic light-by-light
  contribution to $(g-2)_\mu$ with large-$N_c$ Regge models}},  {\em JHEP} {\bf
  03} (2020) 101, [\href{http://arxiv.org/abs/1910.13432}{{\tt
  arXiv:1910.13432}}].

\bibitem{Ludtke:2020moa}
J.~L\"udtke and M.~Procura, {\it {Effects of longitudinal short-distance
  constraints on the hadronic light-by-light contribution to the muon
  ${g-2}$}},  {\em Eur. Phys. J. C} {\bf 80} (2020), no.~12 1108,
  [\href{http://arxiv.org/abs/2006.00007}{{\tt arXiv:2006.00007}}].

\bibitem{Colangelo:2021nkr}
G.~Colangelo, F.~Hagelstein, M.~Hoferichter, L.~Laub, and P.~Stoffer, {\it
  {Short-distance constraints for the longitudinal component of the hadronic
  light-by-light amplitude: an update}},  {\em Eur. Phys. J. C} {\bf 81}
  (2021), no.~8 702, [\href{http://arxiv.org/abs/2106.13222}{{\tt
  arXiv:2106.13222}}].

\bibitem{Knecht:2020xyr}
M.~Knecht, {\it {On some short-distance properties of the fourth-rank hadronic
  vacuum polarization tensor and the anomalous magnetic moment of the muon}},
  {\em JHEP} {\bf 08} (2020) 056, [\href{http://arxiv.org/abs/2005.09929}{{\tt
  arXiv:2005.09929}}].

\bibitem{Bijnens:2021jqo}
J.~Bijnens, N.~Hermansson-Truedsson, L.~Laub, and A.~Rodr\'\i{}guez-S\'anchez,
  {\it {The two-loop perturbative correction to the $(g-2)_\mu$ HLbL at short
  distances}},  {\em JHEP} {\bf 04} (2021) 240,
  [\href{http://arxiv.org/abs/2101.09169}{{\tt arXiv:2101.09169}}].

\bibitem{Bijnens:2022itw}
J.~Bijnens, N.~Hermansson-Truedsson, and A.~Rodr\'\i{}guez-S\'anchez, {\it
  {Constraints on the hadronic light-by-light in the Melnikov-Vainshtein
  regime}},  {\em JHEP} {\bf 02} (2023) 167,
  [\href{http://arxiv.org/abs/2211.17183}{{\tt arXiv:2211.17183}}].

\bibitem{Pauk:2014rta}
V.~Pauk and M.~Vanderhaeghen, {\it {Single meson contributions to the muon's
  anomalous magnetic moment}},  {\em Eur. Phys. J.} {\bf C74} (2014) 3008,
  [\href{http://arxiv.org/abs/1401.0832}{{\tt arXiv:1401.0832}}].

\bibitem{Jegerlehner:2017gek}
F.~Jegerlehner, {\it {The Anomalous Magnetic Moment of the Muon, Second
  Edition}},  {\em Springer Tracts Mod. Phys.} {\bf 274} (2017) pp.1--693.

\bibitem{Dorokhov:2019tjc}
A.~E. Dorokhov, A.~P. Martynenko, F.~A. Martynenko, A.~E. Radzhabov, and A.~S.
  Zhevlakov, {\it {The LbL contribution to the muon g-2 from the axial-vector
  mesons exchanges within the nonlocal quark model}},  {\em EPJ Web Conf.} {\bf
  212} (2019) 05001, [\href{http://arxiv.org/abs/1910.07815}{{\tt
  arXiv:1910.07815}}].

\bibitem{Masjuan:2020jsf}
P.~Masjuan, P.~Roig, and P.~Sanchez-Puertas, {\it {The interplay of transverse
  degrees of freedom and axial-vector mesons with short-distance constraints in
  ${g-2}$}},  {\em J. Phys. G} {\bf 49} (2022), no.~1 015002,
  [\href{http://arxiv.org/abs/2005.11761}{{\tt arXiv:2005.11761}}].

\bibitem{Radzhabov:2023odj}
A.~E. Radzhabov, A.~S. Zhevlakov, A.~P. Martynenko, and F.~A. Martynenko, {\it
  {Light-by-light contribution to the muon anomalous magnetic moment from the
  axial-vector mesons exchanges within the nonlocal quark model}},  {\em Phys.
  Rev. D} {\bf 108} (2023), no.~1 014033,
  [\href{http://arxiv.org/abs/2301.12641}{{\tt arXiv:2301.12641}}].

\bibitem{Leutgeb:2019gbz}
J.~Leutgeb and A.~Rebhan, {\it {Axial vector transition form factors in
  holographic QCD and their contribution to the anomalous magnetic moment of
  the muon}},  {\em Phys. Rev. D} {\bf 101} (2020) 114015,
  [\href{http://arxiv.org/abs/1912.01596}{{\tt arXiv:1912.01596}}].

\bibitem{Cappiello:2019hwh}
L.~Cappiello, O.~Cat\`a, G.~D'Ambrosio, D.~Greynat, and A.~Iyer, {\it
  {Axial-vector and pseudoscalar mesons in the hadronic light-by-light
  contribution to the muon $(g-2)$}},  {\em Phys. Rev. D} {\bf 102} (2020)
  016009, [\href{http://arxiv.org/abs/1912.02779}{{\tt arXiv:1912.02779}}].

\bibitem{Hoferichter:2020lap}
M.~Hoferichter and P.~Stoffer, {\it {Asymptotic behavior of meson transition
  form factors}},  {\em JHEP} {\bf 05} (2020) 159,
  [\href{http://arxiv.org/abs/2004.06127}{{\tt arXiv:2004.06127}}].

\bibitem{Grigoryan:2008up}
H.~R. Grigoryan and A.~V. Radyushkin, {\it {Anomalous Form Factor of the
  Neutral Pion in Extended AdS/QCD Model with Chern-Simons Term}},  {\em Phys.
  Rev.} {\bf D77} (2008) 115024, [\href{http://arxiv.org/abs/0803.1143}{{\tt
  arXiv:0803.1143}}].

\bibitem{Leutgeb:2021mpu}
J.~Leutgeb and A.~Rebhan, {\it {Hadronic light-by-light contribution to the
  muon g-2 from holographic QCD with massive pions}},  {\em Phys. Rev. D} {\bf
  104} (2021), no.~9 094017, [\href{http://arxiv.org/abs/2108.12345}{{\tt
  arXiv:2108.12345}}].

\bibitem{Leutgeb:2022lqw}
J.~Leutgeb, J.~Mager, and A.~Rebhan, {\it {Hadronic light-by-light contribution
  to the muon g-2 from holographic QCD with solved U(1)$_A$ problem}},  {\em
  Phys. Rev. D} {\bf 107} (2023), no.~5 054021,
  [\href{http://arxiv.org/abs/2211.16562}{{\tt arXiv:2211.16562}}].

\bibitem{Katz:2007tf}
E.~Katz and M.~D. Schwartz, {\it {An Eta primer: Solving the U(1) problem with
  AdS/QCD}},  {\em JHEP} {\bf 08} (2007) 077,
  [\href{http://arxiv.org/abs/0705.0534}{{\tt arXiv:0705.0534}}].

\bibitem{Schafer:2007qy}
T.~Sch\"afer, {\it {Euclidean correlation functions in a holographic model of
  QCD}},  {\em Phys. Rev. D} {\bf 77} (2008) 126010,
  [\href{http://arxiv.org/abs/0711.0236}{{\tt arXiv:0711.0236}}].

\bibitem{Hong:2009zw}
D.~K. Hong and D.~Kim, {\it {Pseudo scalar contributions to light-by-light
  correction of muon $g-2$ in AdS/QCD}},  {\em Phys. Lett.} {\bf B680} (2009)
  480--484, [\href{http://arxiv.org/abs/0904.4042}{{\tt arXiv:0904.4042}}].

\bibitem{Casero:2007ae}
R.~Casero, E.~Kiritsis, and A.~Paredes, {\it {Chiral symmetry breaking as open
  string tachyon condensation}},  {\em Nucl. Phys. B} {\bf 787} (2007) 98--134,
  [\href{http://arxiv.org/abs/hep-th/0702155}{{\tt hep-th/0702155}}].

\bibitem{Jarvinen:2022mys}
M.~J\"arvinen, E.~Kiritsis, F.~Nitti, and E.~Pr\'eau, {\it {Tachyon-dependent
  Chern-Simons terms and the V-QCD baryon}},  {\em JHEP} {\bf 12} (2022) 160,
  [\href{http://arxiv.org/abs/2209.05868}{{\tt arXiv:2209.05868}}].

\bibitem{Zanke:2021wiq}
M.~Zanke, M.~Hoferichter, and B.~Kubis, {\it {On the transition form factors of
  the axial-vector resonance $f_1(1285)$ and its decay into $e^+e^-$}},  {\em
  JHEP} {\bf 07} (2021) 106, [\href{http://arxiv.org/abs/2103.09829}{{\tt
  arXiv:2103.09829}}].

\bibitem{Iatrakis:2010zf}
I.~Iatrakis, E.~Kiritsis, and A.~Paredes, {\it {An AdS/QCD model from Sen's
  tachyon action}},  {\em Phys. Rev. D} {\bf 81} (2010) 115004,
  [\href{http://arxiv.org/abs/1003.2377}{{\tt arXiv:1003.2377}}].

\bibitem{Iatrakis:2010jb}
I.~Iatrakis, E.~Kiritsis, and A.~Paredes, {\it {An AdS/QCD model from tachyon
  condensation: II}},  {\em JHEP} {\bf 11} (2010) 123,
  [\href{http://arxiv.org/abs/1010.1364}{{\tt arXiv:1010.1364}}].

\bibitem{Jarvinen:2011qe}
M.~J{\"a}rvinen and E.~Kiritsis, {\it {Holographic Models for QCD in the
  Veneziano Limit}},  {\em JHEP} {\bf 03} (2012) 002,
  [\href{http://arxiv.org/abs/1112.1261}{{\tt arXiv:1112.1261}}].

\bibitem{Cordova:2019jnf}
C.~C\'ordova, D.~S. Freed, H.~T. Lam, and N.~Seiberg, {\it {Anomalies in the
  Space of Coupling Constants and Their Dynamical Applications I}},  {\em
  SciPost Phys.} {\bf 8} (2020), no.~1 001,
  [\href{http://arxiv.org/abs/1905.09315}{{\tt arXiv:1905.09315}}].

\bibitem{Kanno:2021bze}
H.~Kanno and S.~Sugimoto, {\it {Anomaly and superconnection}},  {\em PTEP} {\bf
  2022} (2022), no.~1 013B02, [\href{http://arxiv.org/abs/2106.01591}{{\tt
  arXiv:2106.01591}}].

\bibitem{Kraus:2000nj}
P.~Kraus and F.~Larsen, {\it {Boundary string field theory of the D anti-D
  system}},  {\em Phys. Rev. D} {\bf 63} (2001) 106004,
  [\href{http://arxiv.org/abs/hep-th/0012198}{{\tt hep-th/0012198}}].

\bibitem{Takayanagi:2000rz}
T.~Takayanagi, S.~Terashima, and T.~Uesugi, {\it {Brane - anti-brane action
  from boundary string field theory}},  {\em JHEP} {\bf 03} (2001) 019,
  [\href{http://arxiv.org/abs/hep-th/0012210}{{\tt hep-th/0012210}}].

\bibitem{Quillen:1985vya}
D.~Quillen, {\it {Superconnections and the Chern character}},  {\em Topology}
  {\bf 24} (1985), no.~1 89--95.

\bibitem{Erlich:2005qh}
J.~Erlich, E.~Katz, D.~T. Son, and M.~A. Stephanov, {\it {QCD and a holographic
  model of hadrons}},  {\em Phys. Rev. Lett.} {\bf 95} (2005) 261602,
  [\href{http://arxiv.org/abs/hep-ph/0501128}{{\tt hep-ph/0501128}}].

\bibitem{DaRold:2005mxj}
L.~Da~Rold and A.~Pomarol, {\it {Chiral symmetry breaking from five-dimensional
  spaces}},  {\em Nucl. Phys.} {\bf B721} (2005) 79--97,
  [\href{http://arxiv.org/abs/hep-ph/0501218}{{\tt hep-ph/0501218}}].

\bibitem{Abidin:2009aj}
Z.~Abidin and C.~E. Carlson, {\it {Strange hadrons and kaon-to-pion transition
  form factors from holography}},  {\em Phys. Rev. D} {\bf 80} (2009) 115010,
  [\href{http://arxiv.org/abs/0908.2452}{{\tt arXiv:0908.2452}}].

\bibitem{Leutgeb:2022cvg}
J.~Leutgeb, A.~Rebhan, and M.~Stadlbauer, {\it {Hadronic vacuum polarization
  contribution to the muon g-2 in holographic QCD}},  {\em Phys. Rev. D} {\bf
  105} (2022), no.~9 094032, [\href{http://arxiv.org/abs/2203.16508}{{\tt
  arXiv:2203.16508}}].

\bibitem{Shifman:1978bx}
M.~A. Shifman, A.~I. Vainshtein, and V.~I. Zakharov, {\it {QCD and Resonance
  Physics. Theoretical Foundations}},  {\em Nucl. Phys. B} {\bf 147} (1979)
  385--447.

\bibitem{Melic:2002ij}
B.~Melic, D.~Mueller, and K.~Passek-Kumericki, {\it {Next-to-next-to-leading
  prediction for the photon to pion transition form-factor}},  {\em Phys. Rev.}
  {\bf D68} (2003) 014013, [\href{http://arxiv.org/abs/hep-ph/0212346}{{\tt
  hep-ph/0212346}}].

\bibitem{Hechenberger:2023ljn}
F.~Hechenberger, J.~Leutgeb, and A.~Rebhan, {\it {Radiative meson and glueball
  decays in the Witten-Sakai-Sugimoto model}},  {\em Phys. Rev. D} {\bf 107}
  (2023), no.~11 114020, [\href{http://arxiv.org/abs/2302.13379}{{\tt
  arXiv:2302.13379}}].

\bibitem{Colangelo:2024xfh}
P.~Colangelo, F.~Giannuzzi, and S.~Nicotri, {\it {Hadronic light-by-light
  scattering contributions to $(g-2)_\mu$ from axial-vector and tensor mesons
  in the holographic soft-wall model}},  {\em Phys. Rev. D} {\bf 109} (2024),
  no.~9 094036, [\href{http://arxiv.org/abs/2402.07579}{{\tt
  arXiv:2402.07579}}].

\bibitem{Gursoy:2007cb}
U.~G{\"u}rsoy and E.~Kiritsis, {\it {Exploring improved holographic theories
  for QCD: Part I}},  {\em JHEP} {\bf 02} (2008) 032,
  [\href{http://arxiv.org/abs/0707.1324}{{\tt arXiv:0707.1324}}].

\bibitem{Gursoy:2007er}
U.~G{\"u}rsoy, E.~Kiritsis, and F.~Nitti, {\it {Exploring improved holographic
  theories for QCD: Part II}},  {\em JHEP} {\bf 02} (2008) 019,
  [\href{http://arxiv.org/abs/0707.1349}{{\tt arXiv:0707.1349}}].

\bibitem{Hoferichter:2018kwz}
M.~Hoferichter, B.-L. Hoid, B.~Kubis, S.~Leupold, and S.~P. Schneider, {\it
  {Dispersion relation for hadronic light-by-light scattering: pion pole}},
  {\em JHEP} {\bf 10} (2018) 141, [\href{http://arxiv.org/abs/1808.04823}{{\tt
  arXiv:1808.04823}}].

\bibitem{Gerardin:2019vio}
A.~G\'erardin, H.~B. Meyer, and A.~Nyffeler, {\it {Lattice calculation of the
  pion transition form factor with $N_f=2+1$ Wilson quarks}},  {\em Phys. Rev.
  D} {\bf 100} (2019), no.~3 034520,
  [\href{http://arxiv.org/abs/1903.09471}{{\tt arXiv:1903.09471}}].

\bibitem{Ludtke:2024ase}
J.~L\"udtke, M.~Procura, and P.~Stoffer, {\it {Dispersion relations for the
  hadronic VVA correlator}},  \href{http://arxiv.org/abs/2410.11946}{{\tt
  arXiv:2410.11946}}.

\bibitem{Achard:2001uu}
{\bf L3} Collaboration, P.~Achard et~al., {\it {$f_1(1285)$ formation in
  two-photon collisions at LEP}},  {\em Phys. Lett.} {\bf B526} (2002)
  269--277, [\href{http://arxiv.org/abs/hep-ex/0110073}{{\tt hep-ex/0110073}}].

\bibitem{Achard:2007hm}
{\bf L3} Collaboration, P.~Achard et~al., {\it {Study of resonance formation in
  the mass region 1400 -- 1500 MeV through the reaction $\gamma\gamma \to K^0_S
  K^{\pm} \pi^{\mp}$}},  {\em JHEP} {\bf 03} (2007) 018.

\bibitem{ParticleDataGroup:2024cfk}
{\bf Particle Data Group} Collaboration, S.~Navas et~al., {\it {Review of
  particle physics}},  {\em Phys. Rev. D} {\bf 110} (2024), no.~3 030001.

\bibitem{Hoferichter:2023tgp}
M.~Hoferichter, B.~Kubis, and M.~Zanke, {\it {Axial-vector transition form
  factors and $e^+ e^- \to f_1 \pi^+ \pi^-$}},  {\em JHEP} {\bf 08} (2023) 209,
  [\href{http://arxiv.org/abs/2307.14413}{{\tt arXiv:2307.14413}}].

\bibitem{SND:2019rmq}
{\bf SND} Collaboration, M.~N. Achasov et~al., {\it {Search for direct
  production of the $f_1(1285)$ resonance in $e^+e^-$ collisions}},  {\em Phys.
  Lett. B} {\bf 800} (2020) 135074,
  [\href{http://arxiv.org/abs/1906.03838}{{\tt arXiv:1906.03838}}].

\bibitem{Leutwyler:1997yr}
H.~Leutwyler, {\it {On the 1/N expansion in chiral perturbation theory}},  {\em
  Nucl. Phys. B Proc. Suppl.} {\bf 64} (1998) 223--231,
  [\href{http://arxiv.org/abs/hep-ph/9709408}{{\tt hep-ph/9709408}}].

\bibitem{Escribano:2005qq}
R.~Escribano and J.-M. Fr{\`e}re, {\it {Study of the $\eta$-$\eta'$ system in
  the two mixing angle scheme}},  {\em JHEP} {\bf 06} (2005) 029,
  [\href{http://arxiv.org/abs/hep-ph/0501072}{{\tt hep-ph/0501072}}].

\bibitem{Bali:2021qem}
{\bf RQCD} Collaboration, G.~S. Bali, V.~Braun, S.~Collins, A.~Sch\"afer, and
  J.~Simeth, {\it {Masses and decay constants of the \ensuremath{\eta} and
  \ensuremath{\eta}' mesons from lattice QCD}},  {\em JHEP} {\bf 08} (2021)
  137, [\href{http://arxiv.org/abs/2106.05398}{{\tt arXiv:2106.05398}}].

\bibitem{Roig:2019reh}
P.~Roig and P.~Sanchez-Puertas, {\it {Axial-vector exchange contribution to the
  hadronic light-by-light piece of the muon anomalous magnetic moment}},  {\em
  Phys. Rev. D} {\bf 101} (2020) 074019,
  [\href{http://arxiv.org/abs/1910.02881}{{\tt arXiv:1910.02881}}].

\bibitem{Yang:2007zt}
K.-C. Yang, {\it {Light-cone distribution amplitudes of axial-vector mesons}},
  {\em Nucl. Phys. B} {\bf 776} (2007) 187--257,
  [\href{http://arxiv.org/abs/0705.0692}{{\tt arXiv:0705.0692}}].

\bibitem{Eichmann:2024glq}
G.~Eichmann, C.~S. Fischer, T.~Haeuser, and O.~Regenfelder, {\it {Axial-vector
  and scalar contributions to hadronic light-by-light scattering}},
  \href{http://arxiv.org/abs/2411.05652}{{\tt arXiv:2411.05652}}.

\bibitem{Holz:2024lom}
S.~Holz, M.~Hoferichter, B.-L. Hoid, and B.~Kubis, {\it {A precision evaluation
  of the $\eta$- and $\eta'$-pole contributions to hadronic light-by-light
  scattering in the anomalous magnetic moment of the muon}},
  \href{http://arxiv.org/abs/2411.08098}{{\tt arXiv:2411.08098}}.

\bibitem{Colangelo:2023een}
P.~Colangelo, F.~Giannuzzi, and S.~Nicotri, {\it {$\pi^0$, $\eta$, $\eta'$
  two-photon transition form factors in the holographic soft-wall model and
  contributions to $(g-2)_\mu$}},  {\em Phys. Lett. B} {\bf 840} (2023) 137878,
  [\href{http://arxiv.org/abs/2301.06456}{{\tt arXiv:2301.06456}}].

\bibitem{Mager:2025pvz}
J.~Mager, L.~Cappiello, J.~Leutgeb, and A.~Rebhan, {\it {Longitudinal
  short-distance constraints on hadronic light-by-light scattering and tensor
  meson contributions to the muon $g-2$}},
  \href{http://arxiv.org/abs/2501.19293}{{\tt arXiv:2501.19293}}.

\bibitem{Kwee:2007dd}
H.~J. Kwee and R.~F. Lebed, {\it {Pion form-factors in holographic QCD}},  {\em
  JHEP} {\bf 01} (2008) 027, [\href{http://arxiv.org/abs/0708.4054}{{\tt
  arXiv:0708.4054}}].

\bibitem{Hoferichter:2024vbu}
M.~Hoferichter, P.~Stoffer, and M.~Zillinger, {\it {Complete Dispersive
  Evaluation of the Hadronic Light-by-Light Contribution to Muon g-2}},  {\em
  Phys. Rev. Lett.} {\bf 134} (2025), no.~6 061902,
  [\href{http://arxiv.org/abs/2412.00190}{{\tt arXiv:2412.00190}}].

\bibitem{Hoferichter:2024bae}
M.~Hoferichter, P.~Stoffer, and M.~Zillinger, {\it {Dispersion relation for
  hadronic light-by-light scattering: subleading contributions}},  {\em JHEP}
  {\bf 02} (2025) 121, [\href{http://arxiv.org/abs/2412.00178}{{\tt
  arXiv:2412.00178}}].

\bibitem{Danilkin:2016hnh}
I.~Danilkin and M.~Vanderhaeghen, {\it {Light-by-light scattering sum rules in
  light of new data}},  {\em Phys. Rev.} {\bf D95} (2017), no.~1 014019,
  [\href{http://arxiv.org/abs/1611.04646}{{\tt arXiv:1611.04646}}].

\bibitem{Cappiello:2025fyf}
L.~Cappiello, J.~Leutgeb, J.~Mager, and A.~Rebhan, {\it {Tensor meson
  transition form factors in holographic QCD and the muon $g-2$}},
  \href{http://arxiv.org/abs/2501.09699}{{\tt arXiv:2501.09699}}.

\bibitem{Cappiello:2021vzi}
L.~Cappiello, O.~Cat\`a, and G.~D'Ambrosio, {\it {Scalar resonances in the
  hadronic light-by-light contribution to the muon (g-2)}},  {\em Phys. Rev. D}
  {\bf 105} (2022), no.~5 056020, [\href{http://arxiv.org/abs/2110.05962}{{\tt
  arXiv:2110.05962}}].

\bibitem{Kroll:2016mbt}
P.~Kroll, {\it {A study of the $\gamma ^*$ \textendash{} $f_{0}(980)$
  transition form factors}},  {\em Eur. Phys. J. C} {\bf 77} (2017), no.~2 95,
  [\href{http://arxiv.org/abs/1610.01020}{{\tt arXiv:1610.01020}}].

\bibitem{Shtabovenko:2020gxv}
V.~Shtabovenko, R.~Mertig, and F.~Orellana, {\it {FeynCalc 9.3: New features
  and improvements}},  {\em Comput. Phys. Commun.} {\bf 256} (2020) 107478,
  [\href{http://arxiv.org/abs/2001.04407}{{\tt arXiv:2001.04407}}].

\bibitem{Hahn:1998yk}
T.~Hahn and M.~Perez-Victoria, {\it {Automatized one loop calculations in
  four-dimensions and D-dimensions}},  {\em Comput. Phys. Commun.} {\bf 118}
  (1999) 153--165, [\href{http://arxiv.org/abs/hep-ph/9807565}{{\tt
  hep-ph/9807565}}].

\end{thebibliography}\endgroup

\end{document}